\documentclass[nofootinbib,notitlepage,superscriptaddress]{revtex4-1}
\pdfoutput=1

\usepackage{aas_macros,amsmath,amssymb,esint,graphicx,overpic,xcolor,mathrsfs,stackengine,array,physics}
\usepackage[bottom]{footmisc}
\usepackage[colorlinks=true]{hyperref}
\definecolor{red}{RGB}{228,26,28}
\definecolor{green}{RGB}{77,175,74}
\definecolor{blue}{RGB}{55,126,184}
\definecolor{purple}{RGB}{152,78,163}

\let\originalleft\left
\let\originalright\right
\renewcommand{\left}{\mathopen{}\mathclose\bgroup\originalleft}
\renewcommand{\right}{\aftergroup\egroup\originalright}

\newcommand{\br}[1]{\left[#1\right]}
\newcommand{\cu}[1]{\left\{#1\right\}}
\newcommand{\pa}[1]{\left(#1\right)}
\newcommand{\ed}{\mathop{}\!\mathrm{d}}

\renewcommand{\O}[1]{\mathcal{O}\pa{#1}}

\newcommand{\nhm}{\br{\mathrm{nhm}}}
\newcommand{\near}{\br{\mathrm{near}}}
\newcommand{\ext}{\br{\mathrm{ext}}}
\newcommand{\sub}{\br{\mathrm{sub}}}
\newcommand{\cm}{\mathrm{cm}}

\newcolumntype{P}[1]{>{\raggedright\arraybackslash}p{#1}}

\usepackage{comment}

\begin{document}

\title{Near-Horizon Collisions around Near-Extremal Black Holes}

\author{Delilah E.~A. Gates}
\email{delilah.gates@cfa.harvard.edu}
\affiliation{Center for Astrophysics $\arrowvert$ Harvard \& Smithsonian, 60 Garden Street, Cambridge, MA 02138, USA}
\affiliation{Black Hole Initiative at Harvard University, 20 Garden Street, Cambridge, MA 02138, USA}

\begin{abstract}
Black holes have sometimes been described as astrophysical particle accelerators because 
finite-energy particles can collide near the horizon with divergent center-of-mass (CM) energy. 
The collisions are classified by the radial motion of the constituent particles at the moment of collision, with each class exhibiting a distinct near-horizon behavior. 
Divergence in the CM energy is sourced by the difference in the rate at which the collision radius approaches the horizon and the rate at which a particle's angular momentum is tuned to a critical value set by the superradiant bound.
To produce a high-energy collision around a near-extremal BH, at least one particle must approach criticality slower than the collision radius approaches the horizon. 
When both particles are ingoing or outgoing, it is additionally required that the particles approach criticality at different rates.
Using a novel multi-scaling limit, we calculate the explicit form for the divergent CM energy.
The angular momentum of some circular orbiters---including those at the innermost stable circular orbit radius and innermost bound circular orbit radius---approach criticality as the BH approaches extremality, indicating near-extremal BHs may be a natural environment for these high-energy collisions. 
\end{abstract}
\maketitle

\section{Introduction}
\label{sec:Introduction}

A Kerr black hole (BH) can act as a particle accelerator allowing for collisions of divergent center-of-mass (CM) energy in the vicinity of the horizon. Such high-energy collisions have been shown to occur if one particle has specific 
angular momentum equal to the critical value 
\begin{align}
    \label{eq:CriticalAngularMomentum}
    L_*=\Omega_\mathrm{H}^{-1}\mu
\end{align}
where 
$\mu$ is the specific energy of the particle and 
$\Omega_\mathrm{H}$ is the angular velocity of the horizon, i.e., the particle is at the bound of superradiance. The existence of these high-energy collisions was first identified by Ba\~ndos, Silk, and West, and is thus called the BSW effect \cite{Banados2009}. 

Explicit examples of high-energy collisions between a critical particle and generic (non-critical) particle around a rapidly rotating BH have been worked out for two configurations: a pair of infalling particles colliding at the horizon, and an infalling generic particle colliding with a circular orbiter---wherein the orbiter is critical by virtue of having the angular momentum necessary to maintain its fixed radius close to the BH \cite{Jacobson2010,Grib2010,Grib2011a,Harada2011a,Harada2011b}. These examples exhibit CM energy which diverges as the BH tends to maximal spin, a feature intimately related to the special geometry of the near-extremal BH. In the rapidly rotating regime, the near-horizon region of a BH gets stretched into a throat-like geometry with proper radial depth which diverges as the BH approaches extremality \cite{Bardeen1972}. The throat is described by the Near-Horizon Extremal Kerr (NHEK) metric and is the arena in which the high-energy collisions take place \cite{Bardeen1999}.

In this paper, we take a careful look at near-horizon collisions around a near-extremal Kerr BH using a novel multi-scaling limit that tracks the rates at which we take 1) the collision radius to the horizon, 2) the colliding particles to criticality, and 3) the BH spin to extremality. This allows us to identify conditions under which the CM energy diverges, unifying previous results, and offering new insight into their interpretation. 

The outline is as follows: We start with a review of the CM energy of a pair of particles colliding around a sub-extremal Kerr BH where the near-horizon behavior of the CM energy can be classified by the radial motion of the constituent particles at the moment of collision (Sec.~\ref{sec:CollisionInKerr}). 
Next, we discuss the geometry of a near-extremal BH, develop the aforementioned multi-scaling limit,
determine the conditions which lead to CM energy divergence, and calculate the explicit form of such divergence.
High-energy collisions require at least one particle approaches criticality slower than the collision radius approaches the horizon---generic particles are included in this case---and, for pairs of ingoing or outgoing particles, the particles must approach criticality at different rates
(Sec.~\ref{sec:CollisionAroundNearExtremeBH}).
We then specialize to high-energy collisions that include a particle orbiting at or plunging from a circular orbit to contextualize the previous explicit examples, all of which fall into this category (Sec.~\ref{sec:CircularOrbiterCollisions}). Lastly, we provide a brief discussion of the naturalness and observability of these high-energy collisions (Sec.~\ref{sec:Discussion}).

\section{near-horizon collisions around spinning BHs}
\label{sec:CollisionInKerr}

A rotating BH of mass $M$ and angular momentum $J$ is described by the Kerr metric, which in Boyer-Lindquist (BL) coordinates $x^\mu=\cu{t,r,\phi,\theta}$ is \cite{Kerr1963,Boyer1967}~\footnote{We use natural units $G_N=c=1$.}
\begin{align}
    \label{eq:Kerr}
    ds^2=&-\frac{\Delta}{\Sigma}\left(\ed t-a\sin^2\theta \ed \phi\right)^2+\frac{\Sigma}{\Delta}\ed r^2+\Sigma \ed \theta^2 +\frac{\sin^2\theta}{\Sigma}\br{\pa{r^2+a^2}\ed \phi-a \ed t}^2 ,
\end{align}
where $a=J/M$, $\Delta=r^2-2Mr+a^2$ and $\Sigma=r^2+a^2\cos^2\theta$.
The event horizon is located at radius
\begin{align}
    \label{eq:Horizon}
    r_\mathrm{H}=M+\sqrt{M^2-a^2},
\end{align}
and has angular velocity $\Omega_{\mathrm{H}}=a/(2M r_\mathrm{H})$.

Timelike geodesics of particle motion in the Kerr geometry can be parameterized by four independent conserved quantities: 
the particle mass $m$, 
specific energy (energy per unit mass) $\mu$, 
specific azimuthal angular momentum $L$, and 
specific Carter constant $Q$. 
The conserved quantities are given by~\footnote{Often the four-momentum is expressed in terms of particle energy, angular momentum, Carter constant $\pa{\mathcal{E},\mathcal{L},\mathcal{Q}}=\pa{m \mu, m L,m^2Q}$ which are not rescaled by the particle mass. When expressed as such, the formulae apply also to null geodesics for which the particle mass $m$ is set to zero. Herein, as we consider only collisions of massive particles, we opt for the expression utilizing the specific energy, angular momentum, and Carter constant for the sake of compactness in expressions we derive later in the paper.}
\begin{subequations}
\begin{align}
    m=&\sqrt{-p^\nu p_\nu},\qquad
    \mu=-\frac{p_t}{m},\qquad
     L=\frac{p_\phi}{m},\\
     Q=&\frac{p_\theta^2-a^2(m^2-p_t^2)\cos^2{\theta}+p_\phi^2\cot^2{\theta}}{m^2},\\
        =&\frac{-m^2r^2-\Delta p_r^2-{\pa{a p_t-p_\phi}^2}}{m^2}+\frac{{\pa{r^2+a^2}^2 p_t + a p_\phi}}{\Delta m^2}.
\end{align}
\label{eq:ConservedQuantities}
\end{subequations}
Stationarity and axisymmetry of the metric result in conserved energy and angular momentum, while the Carter constant can be obtained from the separability of the Hamilton-Jacobi equation \cite{Carter:1968rr}.

Inverting Eq.~\eqref{eq:ConservedQuantities}, the particle four-momentum can be written as $p= p_\nu \ed x^{\nu}$,
\begin{align}
    \label{eq:KerrMomentum}
    \frac{p}{m}=-\mu\ed t+ \bar{s}_r\frac{\sqrt{\mathcal{R}}}{\Delta} \ed r + \bar{s}_\theta\sqrt{\Theta} \ed \theta+ L\ed \phi,
\end{align}
for which the angular and radial potentials are
\begin{align}
    \label{eq:AngularPotential}
    \Theta(\theta)=&Q+a^2\pa{\mu^{2}-1}\cos^2{\theta}-L^2\cot^2{\theta}\\
    \label{eq:RadialPotential}
    \mathcal{R}(r)=&P^2\pa{r}-\Delta(r)T\pa{r},
\end{align}
where
\begin{align}
    P(r)&=\mu\pa{r^2+a^2}-aL,\\
    T(r)&=Q+\pa{L-a\mu}^2+r^2.
\end{align} 
and where $\bar{s}_r$ and $\bar{s}_\theta$ are $\pm1$ when the respective potentials are non-zero, and are zero when the respective potentials are zero. Additionally,
\begin{subequations}
\label{eq:FourMomentum}
\begin{align}
    \frac{ p^t}{m}&=\frac{\pa{r^2+a^2}^2\mu-2MarL}{\Delta\Sigma}+\frac{a^2\mu\sin^2{\theta}}{\Sigma},\\
    \frac{p^r}{m}&=\bar{s}_r\frac{\sqrt{\mathcal{R}}}{\Sigma},\\
    \frac{p^\theta}{m}&=\bar{s}_\theta\frac{\sqrt{\Theta}}{\Sigma},\\
    \frac{p^\phi}{m}&=\frac{2Mar\mu}{\Delta\Sigma}+\frac{\pa{\Sigma-2Mr}L}{\Delta\Sigma\sin^2{\theta}}.
\end{align}
\end{subequations}

Let us consider the collision between two timelike geodesic particles. The CM energy of the collision is given in terms of the total momentum of the system $p_\mathrm{tot}^\nu=p_1^\nu+p_2^\nu$ as follows
\begin{align}
    E_{\cm}^2=-p_{\mathrm{tot}} \cdot p_{\mathrm{tot}}=m_1^2+m_2^2-2 p_1\cdot p_2.
\end{align}
This quantity is given by 
\begin{align}
    \frac{E_{\cm}^2}{m_1m_2}=&\frac{m_1^2+m_2^2}{m_1m_2}+\frac{2\pa{\chi-\mathcal{Y}}}{\Sigma},\label{eq:Ecm}\\
    \chi=&\frac{P_1P_2-\bar{s}_{r_1} \bar{s}_{r_2}\sqrt{\mathcal{R}_1}\sqrt{\mathcal{R}_2}}{\Delta},\label{eq:ECMchi}\\
    \mathcal{Y}=&\frac{\pa{L_1-\mu_1 a \sin^2\theta}\pa{L_2-\mu_2 a \sin^2\theta}}{\sin^2\theta} +\bar{s}_{\theta_1} \bar{s}_{\theta_2} \sqrt{\Theta_1}\sqrt{\Theta_2}.
\end{align}
With the exception of the radially dependent term $\chi$, all the terms of the CM energy \eqref{eq:Ecm} are clearly finite for all $r\ge r_\mathrm{H}$. Thus, if $\chi$ \eqref{eq:ECMchi} is divergent in the multi-scaling limit defined in \eqref{eq:ScalingLimit}, the CM energy is
\begin{align}
    \frac{E_{\cm}^2}{m_1m_2}\approx\frac{2\chi}{M^2\pa{1+\cos^2\theta}}.
\end{align}

\begin{figure}
    \centering
    \includegraphics[width=.7\columnwidth]{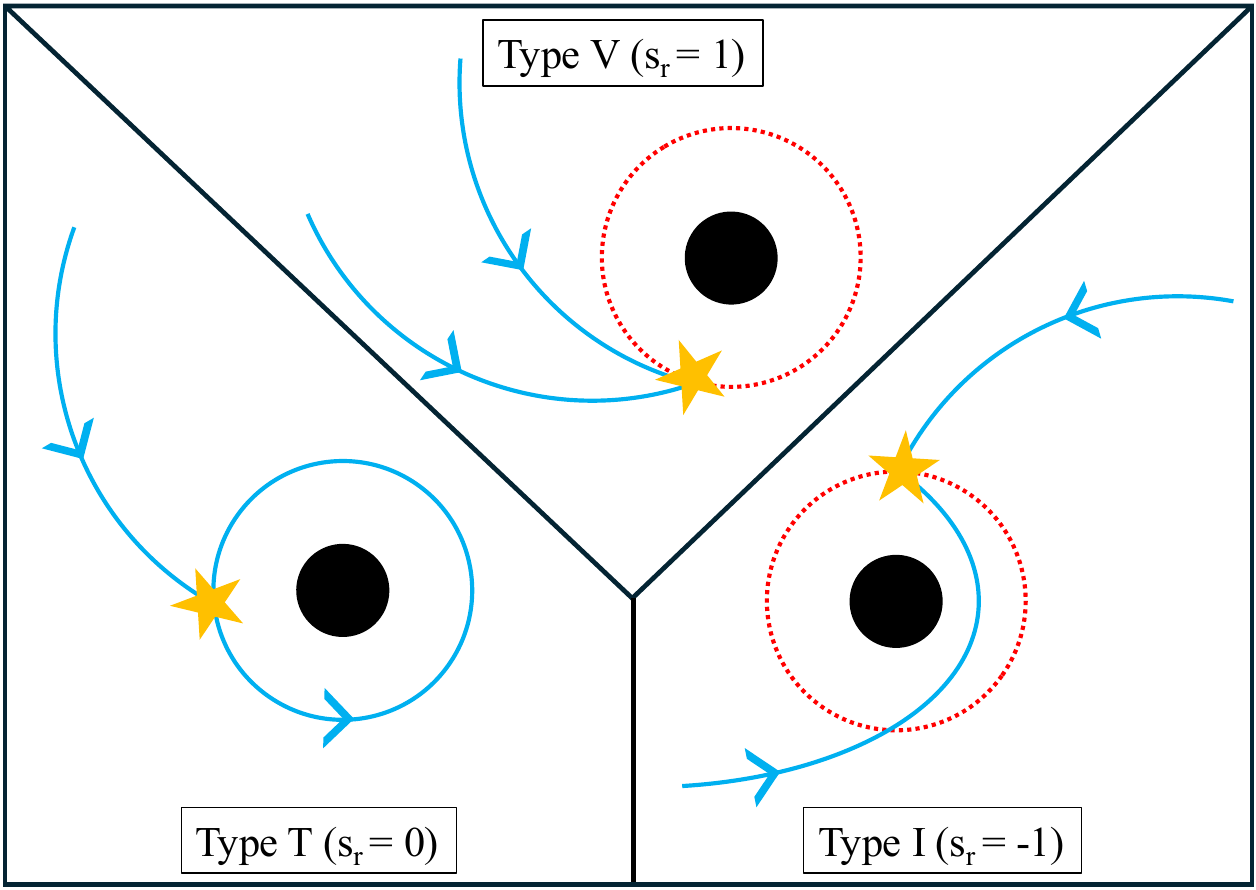}
    \caption{Schematic depictions of particles colliding near a BH. Collisions are categorized by the radial component of particles' momenta at the moment of collision. The particle trajectories are shown as solid blue curves (with arrows for indicating direction), collision location as a yellow star, and the BH horizon as a black circle. Additionally, the collision radius is shown as a dotted red curve in the Type V and I panels.}
	\label{fig:CollisionTypes}
\end{figure}

Now, we will examine the near-horizon limit of $\chi$ \eqref{eq:Ecm} consider the implications for CM energy divergence.
We evaluate the near-horizon limit of $\chi$ in three distinct classes distinguished by the radial motion of the particles at the moment of collision. The classes are Type I in which one particle is outgoing and the other is ingoing, Type T in which one particle has no radial motion, and Type V in which both particles are either ingoing or outgoing (see Fig.~\ref{fig:CollisionTypes}).
In the near-horizon limit $r\to r_\mathrm{H}$,
\begin{align}
    \label{eq:ECMchiNearHorizon}
    \chi=
    \begin{cases}
        \dfrac{T_2\hat L_1}{2\hat L_2}+\dfrac{T_1 \hat L_2}{2\hat L_1}+\O{r-r_{\mathrm{H}}}, & s_{r}=1\\[3ex]
        \dfrac{\pa{1-s_{r}}a^2\hat L_1\hat L_2}{2\sqrt{M^2-a^2}\pa{r-r_\mathrm{H}}}+\O{1}, & s_{r}\leq0
    \end{cases},
\end{align}
where $\hat L_i\equiv L_i-\Omega_\mathrm{H}^{-1}\mu_i$, and $s_{r}\equiv\mathrm{sgn}\pa{\bar{s}_{r_1} \bar{s}_{r_2}}$. 
Necessarily we impose $\mathrm{sgn}(\hat L_1)=\mathrm{sgn}(\hat L_2)$ to maintain positivity~\footnote{Moving forward we will not explicitly mention the conditions needed to maintain positivity of the CM energy.}.

In Type V collisions, deviation of the collision radius from the horizon does not play a direct role the in divergence of the CM energy---since the particles are not a priori at large boost to one another.
By contrast, the CM energy of Types T and Type T collisions diverges near the horizon with the deviation of the collision radius from the horizon, as the particles are at parametrically large boost to one another.

Examining Eq.~\eqref{eq:ECMchiNearHorizon}, it is evident that taking a particle to criticality \eqref{eq:CriticalAngularMomentum} and/or taking the BH to extremality may change the divergence of the near-horizon CM energy.
However, this formula was derived by taking the limit $r\to r_\mathrm{H}$ while holding fixed the particle momenta and the BH spin at unspecified values. Type Type V case of Eq.~\eqref{eq:ECMchiNearHorizon}, which depends on the angular momenta of the particles and the BH spin, is only guaranteed to be applicable to collisions which satisfy the criteria:
\begin{enumerate}
    \item the particles scale to criticality slower than the collision radius scales to the horizon,
    \item the BH approaches extremality slower than the collision radius scales to the horizon.
\end{enumerate}
The Types T and I cases of Eq.~\eqref{eq:ECMchiNearHorizon} depend only on the particles momenta, and so is applicable for all spin as long as criteria 1 is met.

In the next section, we will develop the formalism needed to calculate the CM energy of near-horizon collision in terms of the relative rates at which the particles approach criticality and the collision radius approaches the BH horizon as we spin up the BH.
As the divergence of the CM energy in these near-horizon collision comes from $\chi$ \eqref{eq:ECMchi}, which we can rewrite as
\begin{align}
    \chi&=\tilde P_1\tilde P_2-s_{r}\sqrt{\tilde P_1^2-T_1}\sqrt{\tilde P_2^2-T_2},
\end{align}
where $\tilde{P}=P/\sqrt{\Delta}$,
the (possible) divergence of the CM energy can be diagnosed by examining $\tilde P_i$ and $T_i$ for particles which approach criticality and around a BH that approaches extremality.
(We note the critical angular momentum is related to the remarkable property of BH superradiance whereby an object with $\mu\leq\Omega_\mathrm{H}L$ can scatter off the BH with amplified energy. For an overview of near-horizon high-energy collisions, we refer the reader to the review on BHs as particle accelerators, Ref.~\cite{Harada2014}, and the references therein.) Next we will work out, in detail, the form of the divergent CM energy for collisions which include critical particles around near-extremal BHs.

\section{high-energy collisions around near-extremal BHS}
\label{sec:CollisionAroundNearExtremeBH}

Several explicit examples of high-energy collisions between generic and critical particles around Kerr BHs have been worked out in the limit that the BH approaches extremality. 
Type V collisions at the horizon which diverge as $E_{\cm}\sim\pa{1-a^2/M^2}^{-1/4}$ have been shown to exist numerically by Jacobson and Sotiriou (JS) and analytically by Grib and Pavlov (GP) and by Harada and Kimura (HK) \cite{Jacobson2010,Grib2010,Grib2011a,Harada2011a,Harada2011b}.
High-energy Type T collisions of generic particles and circular orbiters have been shown to exist around near-extremal BHs. These circular orbiter collisions are only possible for a near-extremal BH as prograde circular orbits only become close to the horizon in the near-extremal limit. JS numerically showed that collisions with an orbiter at the prograde innermost bound circular orbit (IBCO) exhibit a $E_{\cm}\sim\pa{1-a^2/M^2}^{-1/4}$ divergence; while, HK analytically showed that collisions with an orbiter at the prograde innermost stable circular orbit (ISCO) exhibit a $E_{\cm}\sim\pa{1-a^2/M^2}^{-1/6}$ divergence \cite{Jacobson2010,Harada2011a,Harada2011b}~\footnote{The divergent Type T and V scenario including IBCO and ISCO particles studied by JS, GP, and HK have been extended to generalized near-extremal stationary, axisymmetric BHs \cite{Zaslavskii2012b}. This suggests the subsequent results we derive herein for the near-extremal Kerr BH may also extend.}. 
 
While the near-extremal ($a\to M$) case (i.e., BH spin parametrically close to maximal) is the main subject of this work, near-horizon high-energy collisions have been shown to exist around extremal ($a=M$) and sub-extremal ($a<M$) cases. 
Indeed, the divergent collisions between a generic and critical particle were first discovered by BSW in the extremal BH case, which introduced the idea of BHs as astrophysical particle accelerators. BSW examined a pair of ingoing particles originating far from the BH, traveling along geodesics, that collided at the horizon, the so-called ``direct collision scenario'' \cite{Banados2009}. 
Subsequently, Grib and Pavlov (GP) showed the existence of near-horizon high-energy collisions for a sub-extremal BH ($a<M$). At sub-extremality, a critical particle starting at infinity cannot reach the BH horizon along a geodesic. As such high-energy collisions require a ``multiple scattering'' scenario in which one particle has its angular momentum critically tuned through other means, e.g., interacting with other particles in an accretion disk as it migrates towards the horizon or originating from a prior near-horizon scattering process \cite{Grib2010,Grib2011a}~\footnote{Particles with vanishing specific energy can also participate in divergent near-horizon CM energy collisions\cite{Grib2017}.}${}^,$\footnote{The existence of near-horizon high-energy collisions of generic and critical particles has also been shown for the Kerr-Newman BH \cite{Wei2010} as well as general stationary, axisymmetric BHs \cite{Zaslavskii2010,Zaslavskii2012a}.}. We treat the extremal and sub-extremal cases in Apps.~\ref{app:EKCMEnergy} and \ref{app:SubEKCMEnergy}, respectively. 

In this section, we develop a unified framework that organizes all known examples of high-energy collisions. Fully resolving the behavior of near-horizon collisions around a near-extremal BH necessitates keeping track of the rates at which the collision radius approaches the horizon, the particles approach criticality (if they approach it at all), and the BH approaches extremality. The multi-scaling limit required to carry out this calculation is intimately linked to the geometry of the near-extremal BH.

\subsection{Near-horizon, near-critical scaling limit}
\label{sec:NearHorizonCriticalLimit}

BH spin can be defined in terms of a parameter $\kappa=\sqrt{1-a^2/M^2}$ which describes deviation from maximal spin, and which sets the horizon at 
\begin{align}
   r_\mathrm{H}=M\pa{1+\kappa}.
\end{align}
A near-extremal BH can be described by $\kappa\ll 1$. At extremality ($\kappa=0$), the horizon goes to $r=M$. 
However, in the extremal limit, several physically distinct radii such as the prograde photon circular orbit (PCO), IBCO, and ISCO radii also approach $r=M$. In actuality, as the BH spin is increased, the near-horizon region develops a throat-like geometry with proper radial depth that diverges with deviation from extremality $\kappa$. For this reason it is well known that BL coordinates do not resolve the near-horizon physics of near-extremal BHs \cite{Bardeen1972}. 

To resolve the near-horizon physics, we examine the scaling rate of different BL radii near extremality. Radii which scale as
\begin{align}
    \label{eq:RadiusScaling}
    r-M\sim \O{\kappa^p}, \quad 0<p\leq1,
\end{align}
in the extremal limit approach the horizon.
Radii which scale with the same value of $p$ maintain finite proper distance from one another in the extremal limit; radii which scale with different values of $p$ have divergent proper distance in the extremal limit. 
All BL radii which scale with $0<p<1$ are described by the limiting metric known as the near-horizon extreme Kerr (NHEK) metric. Radii which scale with $p=1$ are described by a different limiting metric, the near-NHEK metric \cite{Bardeen1999,Bredberg2010} (see App.~\ref{app:NHEK}).
On the other hand, radii that do not scale to $r=M$ (e.g., the retrograde PCO, IBCO, and ISCO radii) remain in the far region of the geometry, which is described by the extreme Kerr metric.

\begin{figure}
    \centering
	\includegraphics[width=.7\columnwidth]{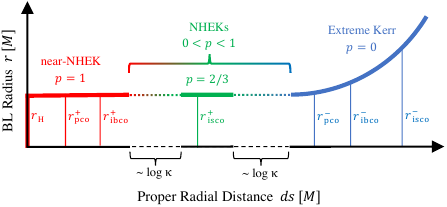}
	\caption{Geometry of a near-extremal BH. In the near-extremal limit $\kappa=\sqrt{1-a^2/M^2}\ll 1$, the Kerr BH develops a throat-like geometry of divergent proper depth. All radii which approach $r=M$ at the same rate ($r-M\sim \kappa^p$) as the BH approaches extremality are described by the same limiting metric (near-NHEK, NHEK, or extreme Kerr). The horizon and the prograde and retrograde PCO, IBCO, and ISCO radii are marked.}
	\label{fig:NearExtremeGeometry}
\end{figure}

Each set of radii with scaling $r-M\sim\kappa^p$ is physically distinct. Radii which scale with the same power $p$ end up in the same NHEK and are said to exist in the same ``$p$-band'' of the near-extremal BH geometry. The horizon and prograde PCO and IBCO radii belong to the $p=1$ band (i.e., the near-NHEK) while the prograde ISCO radius belongs to the $p=2/3$ band. The far region can be thought of as the $p=0$ band. The full geometry of the near-extremal BH can be thought of as the ensemble of all the infinitely many $p$-bands (see Fig.~\ref{fig:NearExtremeGeometry}). Any process which involves both the rate at which a radius approaches the horizon $\kappa^p$ and the rate at which the BH spin increases $\kappa$ must be calculated as a limit applied to the sub-extremal case \cite{Gates2020}.

We must also consider the rates at which particles approach criticality. As the BH spin approaches maximum, the critical angular momentum \eqref{eq:CriticalAngularMomentum} behaves as 
\begin{align}
    L_*=\Omega_\mathrm{H}^{-1}\mu=2M\mu \br{1+\kappa+\O{\kappa^2}}, 
\end{align}
which reaches $L=2M\mu$ at extremality. 
Particles whose angular momenta scale as
\begin{align}
    L-2M\mu\sim\O{\kappa^q},\quad 0<q,
    \label{eq:AngularMomentumScaling}
\end{align}
in the extremal limit approach criticality; these include prograde IBCO and ISCO orbiters for which $q= 1,\ 4/3$, respectively.
Generic particles, those with angular momenta which does not scale to criticality in the near extremal limit, include retrograde PCO, ISCO, and IBCO orbiters.

Recalling that that the CM energy or near-horizon collisions may diverge according to relative rates at which the collision radius approaches the horizon and the particle angular momenta approach criticality, we need to resolve $r-r_{\mathrm H}$, and $L-\Omega_\mathrm{H}^{-1}\mu$ to leading order in small parameter $\kappa$.
With respect to $\kappa$, a particle's angular momentum may be
\begin{itemize}
    \item \emph{generic}: remaining non-critical as the BH approaches extremality,
    \item \emph{near-critical}: approach criticality as the BH approaches extremality,
    \item \emph{precisely critical}: a priori set to criticality before spinning up the BH~\footnote{Future-directed particles have $\mu>-2MaLr/\Pi$, where $\Pi=(r^2+a^2)^2-\Delta a^2\sin^2\theta$. As such, $\mu=L=0$ is not admissible.}.
\end{itemize}
In order to calculate the CM energy of a near-horizon collision to leading order in $\kappa$, we need to track the following
\begin{itemize}
    \item the collision radius to subleading order,
    \item the particle angular momenta to
    \begin{itemize}
        \item leading order when generic/precisely critical,
        \item subleading order when near-critical.
    \end{itemize}
\end{itemize}
That is to say, all near-horizon collisions for which the collision radii and particle angular momenta match to the orders outlined above share the same leading order CM energy behavior in the $\kappa\to0$ limit. 

Motivated by the form of Eqs.~\eqref{eq:RadiusScaling} and \eqref{eq:AngularMomentumScaling}, we will adopt the ansatz
\begin{subequations}
\label{eq:ScalingLimitAnsatz}
\begin{alignat}{2}
    \label{eq:SpinScaling}
    a&=M\sqrt{1-\kappa^2},  &\quad & 0<\kappa\ll1, \\
    \label{eq:RScaling}
    r&=M\pa{1+R\kappa^{p_c}}, &\quad & 0<p_c\leq1, \\
    \label{eq:LScaling}
    L_i&=2M(\mu_i-l_i\kappa^{q_i}), &\quad & 0\le q_i.
\end{alignat}
\end{subequations}

The radial parameter $R$ and angular momentum parameter $l$ are dimensionless. 
We use $p_c$ to control the rate at which the BL collision radius scales to $M$, and $q$ to control the rate at which the angular momentum of a particle scales to $2M\mu$.
Note, we also want to consider cases where one or both particles do not approach criticality as $\kappa\to0$, thus we allow for $q_i=0$ in Eq.~\eqref{eq:LScaling}.
This enables us to describe the three aforementioned particles types: generic particles ($l_i\neq0$ and $q_i=0$), near-critical particles ($l_i\neq 0$ and $q_i>0$), and precisely critical particles ($l_i=0$).

Now, using this ansatz, we can define the near-extremal, near-critical, near-horizon multi-scaling limit of a quantity $f$ as
\begin{align}
    \label{eq:ScalingLimit}
    f(a,r,L_1,L_2)\Big\vert_{\text{Eq.\eqref{eq:ScalingLimitAnsatz}}} \overset{\kappa \to 0}{\sim} \ f_{\near}(\kappa).
\end{align}
For convenience we will also define the dimensionless Carter constant
\begin{align}
    \eta_i=\frac{Q_i}{M^2}.
\end{align}
In the next section we will use this multi-scaling limit to determine the conditions for and the form of divergent CM energy collisions.

\subsection{High-energy collisions}
\label{sec:HighEnergyCollisions}

We diagnose the (possible) divergence of the CM energy by examining $\tilde P$ and $T$ in the multi-scaling limit \eqref{eq:ScalingLimit}. 
In the multi-scaling limit, 
\begin{subequations}
\begin{align}
    \tilde P_{i\near}&=\frac{2M\mathcal{A}}{\sqrt{R^2 -\delta_{p_c,1}}},\\
    \mathcal{A}&
    =\begin{cases}
        \dfrac{l_i}{\kappa^{p_c-q_i}}, &q_i<p_c, l_i\neq0\\[3ex]
        R\mu_i+l_i, & q_i=p_c, l_i\neq0\\[1ex]
        R\mu_i, & (q_i> p_c, l_i\neq 0) \text{ or }(l_i=0)
    \end{cases},
\end{align}
\label{eq:TildesPnear}
\end{subequations}
where $\delta_{x,y}$ is the Kronecker delta function, and
\begin{align}
    \label{eq:Tnear}
    {T_{i\near}}&=
    {\tilde T_{i\near}} + 4 M^2l_i\pa{l_i-\mu_i}\delta_{q_i,0},\\[1ex]
    \label{eq:TildeTnear}
    \tilde{T}_{i\near}&=M^2\pa{1+\eta_i+\mu_i^{2}}.
\end{align}
As $\tilde{P}_{i\near}$ is non-vanishing and $T_{i\near}$ is non-vanishing and finite, $R_i\propto\tilde P_i^2-T_i\geq0$ imposes no constraint on $q_i$ (in contrast to the sub-extremal case).

Divergent CM energy requires that at least one particle has angular momentum which scales to criticality slower than the collision radius scales to the horizon,
$$\text{divergence condition:} \ l_i\neq0,\ q_i<p_c .$$ 
As divergence requires a particle to \emph{scale to criticality}, a precisely critical particle does not satisfy this condition. When only particle 1 satisfies this condition and particle 2 does not, $\chi$ \eqref{eq:ECMchi} in the multi-scaling limit becomes
\begin{align}
    \label{eq:SinglySourced}
    \chi_{\near}&=\tilde P_{1\near}\tilde P_{2\near}-s_{r}\sqrt{\tilde P_{1\near}^2}\sqrt{\tilde P_{2\near}^2-\tilde T_{2\near}}\propto \frac{1}{\kappa^{p_c-q_1}}.
\end{align}
We call this ``singly sourced divergence'' as only one particle participates the form of the CM energy divergence. 
When both particles satisfy the divergence condition with $q_1\leq q_2$, for Types T and I collisions, $\chi$ becomes
\begin{align}
    \label{eq:DoublySourcedTypeIandT}
    \chi_{\near}=(1-s_{r})\tilde P_{1\near}\tilde P_{2\near} \propto \frac{1}{\kappa^{2p_c-q_2-q_1}}.
\end{align}
For Type V collision, CM energy divergence additionally requires $q_1<q_2$, and $\chi$ becomes
\begin{align}
    \label{eq:DoublySourcedTypeV}
    \chi_{\near}=\frac{\tilde T_{2\near} {\tilde P_{1\near}} }{2{\tilde P_{2\near}}} \propto \frac{1}{\kappa^{q_2-q_1}},
\end{align}
We call these ``doubly sourced divergence'' as both particle contribute to the form of the CM energy divergence. 

Let us compare $\chi_{\near}$ derived in this section to the near-horizon limit of $\chi$ away from extremality and criticality found in Eq.~\eqref{eq:ECMchiNearHorizon}. Notice that the divergent Type V collision seen here in Eq.~\eqref{eq:DoublySourcedTypeV} is expected from Eq.~\eqref{eq:ECMchiNearHorizon}, matching if we substitute in $L_i$ of the form \eqref{eq:LScaling}. However, the singly divergent collision Eq.~\eqref{eq:SinglySourced} here cannot be deduced from Eq.~\eqref{eq:ECMchiNearHorizon} because particle 2 approaches criticality faster than the collision radius approaches the horizon. Similarly, the doubly divergent Types T and I collisions Eq.~\eqref{eq:DoublySourcedTypeIandT} here cannot be deduced from Eq.~\eqref{eq:ECMchiNearHorizon} because the BH spin approaches extremality at a rate equal to or faster than the rate at which the collision radius approaches the horizon under the multi-scaling limit \eqref{eq:ScalingLimit}.

Further, we can bound the divergence rates of high-energy collisions. Singly sourced divergence \eqref{eq:SinglySourced} is strongest when the collision takes place in the near-NHEK region ($p_c=1$) and particle 1 is generic or precisely critical, 
\begin{align}
    E_{\cm\near}\propto \kappa^{-(p_c-q_1)/2}\leq\pa{1-a^2/M^2}^{-1/4}.
\end{align}
The doubly sourced divergent Type V collision \eqref{eq:DoublySourcedTypeV} is subject to a similar constraint, 
\begin{align}
    E_{\cm\near}\propto \kappa^{-(q_2-q_1)/2}<\pa{1-a^2/M^2}^{-1/4}.
\end{align}
On the other hand, the doubly sourced divergent Types T and I collision \eqref{eq:DoublySourcedTypeIandT} diverges as
\begin{align}
    E_{\cm\near}\propto\kappa^{-p_c+(q_2+q_1)/2}<\pa{1-a^2/M^2}^{-1/2}.
\end{align}

Summarizing, high-energy collisions require that at least one particle have angular momentum which scales to criticality sufficiently ``slowly'' compared to the rate the collision radius scales to the horizon, allowing for both singly and doubly divergent collisions. Type V collisions additionally require that the particles approach criticality at different rates. Secondly, the CM energy divergence of the singly divergent collisions of all Types and doubly divergent Types T and I collisions is set by the difference in rate at which the collision radius approaches the horizon and the rates at which particles approach criticality. The CM energy divergence of the doubly divergent Type V collision is set by the difference in rates at which the particles approach criticality. The insights in the divergence of various collision types described here carry over to the extremal and sub-extremal cases (see Apps.~\ref{app:EKCMEnergy} and \ref{app:SubEKCMEnergy} for details).

\section{high-energy collisions with near-horizon circular orbit particles}
\label{sec:CircularOrbiterCollisions}

Several configurations leading to divergent CM energy collisions around near-extremal BHs have been explored in previous works. Explicit examples have considered collisions between generic and critical particles confined to the equatorial plane ($\theta=\pi/2$, $p_\theta=0$, $\eta=0$). In these examples, the critical particle is a circular orbiter (i.e., a particle at radius $r=r_\circ$ with specific energy $\mu=\mu_\circ(r_\circ)$ and angular momentum $L=L_\circ(r_\circ)$ required to maintain fixed radius) or a plunging particle (i.e., a particle with circular orbiter specific energy $\mu=\mu_\circ(r_\circ)$ and angular momentum $L=L_\circ(r_\circ)$ at radii $r<r_\circ$). And, the generic particle is either a circular orbit particle or marginally bound zero angular momentum particle (MBZAM), i.e., $\mu=1$ and $L=0$. The collisions have been shown to take place at the horizon and the prograde IBCO and ISCO radii.

\subsection{Circular orbit particles in the near-horizon, near-critical limit}
\label{sec:NearCriticalCircularOrbits}

Massive circular orbiters in the Kerr geometry \eqref{eq:Kerr}, defined by $\mathcal{R}(r_\circ)=\mathcal{R}'(r_\circ)=0$, have specific energy and angular momentum \cite{Bardeen1972}
\begin{subequations}
\label{eq:CircularOrbiter}
\begin{align}
	\mu_\circ^\pm&=\frac{r_\circ^{3/2}-2M\sqrt{r_\circ}\pm a\sqrt{M}}{\sqrt{r_\circ^3-3Mr_\circ^2\pm2a\sqrt{M}r_\circ^{3/2}}},\\
	L_\circ^\pm&=\pm\frac{{\sqrt{M}\pa{r_\circ^2+a^2}\mp2aM\sqrt{r_\circ}}}{\sqrt{r_\circ^3-3Mr_\circ^2\pm2a\sqrt{M}r_\circ^{3/2}}},
\end{align}
\end{subequations}
with prograde/retrograde denoted by the $\pm$. These orbits exist outside of the PCO radius
\begin{align}
    r_{\text{pco}}^\pm=2M\br{1+\cos\br{\frac{2}{3}\arccos\pa{\mp\frac{a}{M}}}},
\end{align}
where the denominator of the specific energy goes to zero.
Bound ($\mu\leq1$) circular orbiters exist at radii down to the IBCO$^\pm$ radius
\begin{align}
    \label{eq:IBCO}
    r_{\mathrm{ibco}}^\pm=2M\mp a +2\sqrt{M\pa{M\mp a}},
\end{align}
where the angular momentum reduces to the simple form $L_{\mathrm{ibco}}^\pm=\pm \br{M+\sqrt{M(M\mp a)}}$. Circular orbits are stable $\mathcal{R}''(r_\circ)<0$ down to the ISCO$^\pm$ radius
\begin{subequations}
\begin{align}
    \label{eq:ISCO}
    r_{\mathrm{isco}}^\pm=&M\br{3+Z_2\mp\sqrt{\pa{3-Z_1}\pa{3+Z_1+2Z_2}}},\\
    Z_1=&1+\sqrt[3]{1-\frac{a^2}{M^2}}\pa{\sqrt[3]{1+\frac{a}{M}}+\sqrt[3]{1-\frac{a}{M}}},\\
	Z_2=&\sqrt{3\frac{a^2}{M^2}+Z_1^2}.
\end{align}
\end{subequations}
Outside the equatorial plane there are spherical orbits of fixed radii which are stable down to the last stable spherical orbit (LSSO), and which are bound down to the last spherical bound orbit (LBSO). Although no closed-form description for these non-equatorial spherical orbits exists \cite{Hod2017,Stein2020,Teo2021}. In the extremal limit, LBSO and LSSO radii which scale to $r=M$ exist only within latitudes $\pm\arccos(2/3)$ and $\pm\arccos(\sqrt{3}-1)$, respectively; and thus, high-energy collisions which include LSSO and LBSO particles exist only within the respective latitude bands \cite{Harada2011b,Hod2017}.

We will consider collisions which include particles on circular orbits and the associated particles which have the same conserved quantities as a circular orbiter and plunge towards the horizon. A plunging particle associated with a stable circular orbiter ($\mu=\mu_\circ(r_\circ), L=L_\circ(r_\circ), r_\circ>r_\mathrm{isco}$) cannot be thought of as plunging \emph{from} the circular orbiter radius $r_\circ$; instead, it travels along a geodesic which begins and ends on the horizon and has a radial turning point strictly less than $r_\circ$. By contrast, a plunger associated with the ISCO (which is marginally stable) or an unstable circular orbiter ($r_\circ \leq r_\mathrm{isco}$) travels along a geodesic which asymptotically spirals out of the circular orbit radius and plunges into the horizon. These plungers can be thought of as plunging from the circular orbit radius \cite{Compere2021}. Hence, we will consider only the particles with ($\mu=\mu_\circ(r_\circ), L=L_\circ(r_\circ), r_\circ\leq r_\mathrm{isco}$) and call them \emph{circular orbit plungers}.

Additionally, we note that a particle with the conserved quantities of an unstable circular orbiter can asymptotically spiral onto or out of the circular orbit radius \cite{Levin2009, Compere2021}. Previous works by JS and HK (Refs. \cite{Jacobson2010,Harada2011a,Harada2011b}) consider collisions between a generic particles and an infalling near-critical particles, with $\mu=1$ and $L_
\mathrm{ibco}^+$, that occurs on the IBCO$^+$ radius. As the IBCO$^+$ orbiter is the limit for such a particle approaching the IBCO radius in these collisions, they are effectively Type T collisions with an IBCO$^+$ orbiter. 

To calculate the CM energy of a high-energy collision, we need the radial parameter $R$ and scaling value $p$ of the collision radius. By \eqref{eq:RScaling}, the near-extremal limit $R\kappa^p \approx r-M$ provides the parameters. These parameters of the ISCO$^\pm$ radii, IBCO$^\pm$ radii, and horizon are provided in Tab.~\ref{tab:NearHorizonNearCriticalParams} (left side). As all retrograde circular orbit radii remain in the asymptotically far region of the near-extremal BH geometry $p=0$, high-energy collisions cannot occur at the retrograde circular orbit radii. However, the IBCO$^+$ radius exists in the $p=1$ band, so prograde circular orbiters can exist in all $p$-bands, and high-energy collisions can occur at any prograde circular orbit radii in the throat.

We also need the specific energy $\mu$, angular momentum parameter $l$, and scaling value $q$ of each participating particle. 
By \eqref{eq:LScaling}, the near-extremal limit $l\kappa^q\approx\mu-L/(2M)$ provides the latter of these. Both ISCO$^+$ and IBCO$^+$ orbiters have angular momenta tuned to criticality.
The specific energy, angular parameter $l$, and scaling value $q$ for IBCO$^\pm$, ISCO$^\pm$ orbiters and MBZAMs are shown in Tab.~\ref{tab:NearHorizonNearCriticalParams} (right side). 
\begin{table}
\centering
 \begin{tabular}{c}
 \begin{tabular}{|c|| c| c| c|} 
 \hline
 Radius  & $R$ & $p$ \\[1ex] 
 \hline
 \hline
 ISCO$^+$ & $2^{1/3}$ & $\frac{2}{3}$ \\[1ex] 
 \hline
 IBCO$^+$ & $\sqrt{2}$ & $1$ \\[1ex] 
 \hline
 ISCO$^-$ & $8$ & $0$ \\[1ex] 
 \hline
 IBCO$^-$ & $2\pa{1+\sqrt{2}}$ & $0$ \\[1ex] 
 \hline
 horizon & $1$ & $1$ \\[1ex] 
 \hline
 \end{tabular}
 \quad
 \begin{tabular}{|c|| c| c| c|} 
 \hline
 Particle Type  & $\mu$ & $l/\mu$ & $q$ \\[1ex] 
 \hline
 \hline
 ISCO$^+$ orbiter& $\frac{1}{\sqrt{3}}$ & $-\frac{3}{2^{4/3}}$ & $\frac{4}{3}$ \\[1ex] 
 \hline
 IBCO$^+$ orbiter & $1$ & $-\frac{1}{\sqrt{2}}$ & $1$ \\[1ex] 
 \hline
 ISCO$^-$ orbiter & $\frac{5}{3^{3/2}}$ & $\frac{16}{5}$ & $0$ \\[1ex] 
 \hline
 IBCO$^-$ orbiter & $1$ & $2+\sqrt{2}$ & $0$ \\[1ex] 
 \hline
 MBZAM & $1$ & $1$ & $0$ \\[1ex] 
 \hline
\end{tabular}
\end{tabular}
\caption{{\bf Left:} radial parameter $R$ and scaling value $p$ of various BL radii. {\bf Right:} specific energy $\mu$, angular momentum parameter $l$, and criticality scaling value $q$ of various particles.}
\label{tab:NearHorizonNearCriticalParams}
\end{table}

We can also consider a circular orbiter \eqref{eq:CircularOrbiter} at a general near-horizon radius $r_\circ=M\pa{1+R_\circ\kappa^{p_\circ}}$ with $0<p_\circ\leq1$. In the $\kappa\to0$ limit, the circular orbiter has specific energy
\begin{align}
    \mu_\circ&\approx
    \begin{cases}
        \dfrac{R_\circ}{\sqrt{3R_\circ^2-4}}, & p_\circ=1\\[3ex]
        \dfrac{1}{\sqrt{3}}, & 0<p_\circ<1
    \end{cases},
\end{align}
and angular momentum parameter and criticality scaling value
\begin{align}
    \label{eq:kappaCircNearHorizon}
    q_\circ&=\mathrm{min}\pa{2p_\circ,2-p_\circ},\\
    \label{eq:lCircNearHorizon}
	l_\circ&=\mu_{\circ}
    \begin{cases}
        -\dfrac{R_\circ^2}{4}, & 0<p_\circ<\dfrac{2}{3}\\[3ex]
        -\dfrac{R_\circ^2}{4}-\dfrac{1}{R_\circ}, & p_\circ=\dfrac{2}{3}\\[3ex]
        -\dfrac{1}{R_\circ}, & \dfrac{2}{3}<p_\circ\le1
    \end{cases}.
\end{align}
As $q_\circ>0$, near-horizon circular orbiters are naturally tuned to criticality (near-critical particles) and circular orbit plungers can source divergence in high-energy collisions. We also note the scaling parameters \eqref{eq:kappaCircNearHorizon} change behavior in the ISCO $p$-band.

\begin{figure*}
    \centering
    \includegraphics[width=\linewidth]{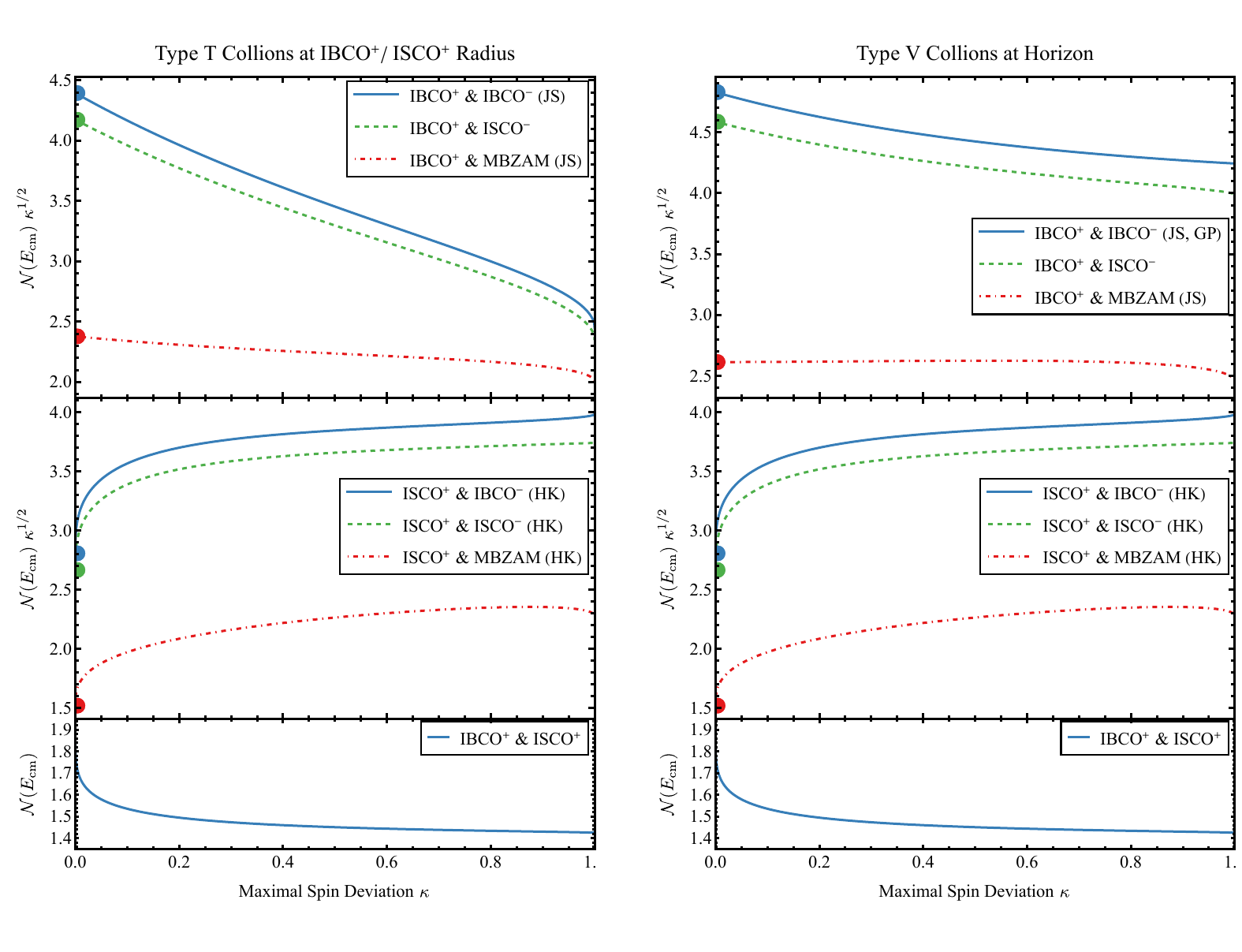}
    \caption{
    CM energies of collisions including prograde IBCO or ISCO particles (orbiters and plungers). We plot the a normalized CM energy $\mathcal{N}(E_{\cm})$ \eqref{eq:NCMEF} as a function of the maximal spin deviation parameter $\kappa=\sqrt{1-a^2/M^2}$. The normalized CM energy has been multiplied by the appropriate powers of $\kappa$ which regularize the CM energy in the near-extremal limit $\kappa\to0$. The left column shows Type T collisions including particles orbiting at the prograde IBCO or ISCO radius. The right column shows Type V collisions occurring at the horizon that include particles plunging from the prograde IBCO/ISCO plungers. We label the prograde IBCO and ISCO particles with ``IBCO$^+$'' and ``ISCO$^+$'', respectively. \\[1ex]
    \textbf{Top panels:} The prograde IBCO orbiter/plunger (left/right) collides with generic particles: retrograde IBCO plunger (``IBCO$^-$''), ISCO plunger (``ISCO$^-$''), and a marginally bound zero angular momentum particle (``MBZAM'').\\ 
    \textbf{Middle panels:} The prograde ISCO orbiter/plunger collides with generic particles.\\
    \textbf{Bottom panels:} The prograde IBCO orbiter/plunger collides with prograde ISCO plunger.\\[1ex]
    The Type T collision of a near-horizon circular orbiter and a generic particle has CM energy which diverges in the near-extremal limit $\kappa\to0$ (details in Sec.~\ref{sec:TypeTCollisions}), as does the Type V collision of a near-horizon circular orbiter plunger and a generic particle at the horizon (details in Sec.~\ref{sec:TypeVCollisions}). In the near-extremal limit, the CM energies of these collisions diverge as $E_{\cm}\sim\mathcal{O} (\kappa^{-p_c/2})$ where $p_c$ is the rate at which the collision radius scales to $r=M$ as $\kappa\to0$ \eqref{eq:RScaling}; $p_c=2/3$ for the ISCO$^+$ radius and $p_c=1$ for the horizon. The analytic formulae for the high-energy CM energy $E_{\cm \near}/\sqrt{m_1m_2}$ for the Type T and V collisions are given by Eqs.~\eqref{eq:TypeTCircOrbs} and \eqref{eq:TypeVCircOrbs}, respectively, and are illustrated as dots along the $\kappa=0$ axis in the top and middle plots. The CM energy of Type T and V collisions of prograde IBCO/ISCO orbiters and plungers colliding with precisely critical particles is finite in the near-extremal limit, as is the CM energy of collisions of prograde IBCO orbiters/plungers colliding with prograde IBCO plungers.
    }
    \label{fig:Collisions}
\end{figure*}

Near-horizon circular orbiter particles (orbiters and plungers) are near-critical particles ($l_i\neq 0,\ q_i>0$) with $p_\circ\geq p_c$. For Type T collisions including a near-horizon circular orbiter, $q_\circ\geq p_\circ=p_c$ by Eq.~\eqref{eq:kappaCircNearHorizon}; so, the circular orbiter does not meet the divergence criteria. 
It is important to note, the critical tuning of the near-horizon circular orbiter is not the cause of the CM energy divergence for these near-horizon Type T collision as necessarily $q_\circ\geq p_c$. Instead, the large boost factor of an incoming particle relative to the near-horizon circular orbiter causes the CM energy divergence.
More generally, we note that the critical tuning of the circular orbit particle (orbiter or plunger) does not contribute to the divergence of the CM energy for a near-horizon collision in its own $p$-band.

A near-horizon circular orbit plunger meets the divergence condition only if $q_\circ<p_c\leq 1$, which by \eqref{eq:kappaCircNearHorizon} occurs for $p_\circ<1/2$. 
However, near-horizon circular orbit plungers, which are restricted to occurring for circular orbits with radii $r_\circ\leq r_\mathrm{isco}^+$, exhibit $p_\circ\geq2/3$ and $q_\circ=2-p_\circ$ according to Eq.~\eqref{eq:kappaCircNearHorizon}. As such, $q_\circ>p_\circ\geq p_c$; so, the circular orbit plunger does not meet the divergence criteria. Thus, a collision including a near-horizon circular orbit particle must have finite or singly divergent CM energy.
(As near-horizon circular orbit particles---specifically the exemplar ISCO$^+$ and IBCO$^+$ orbiters/plungers used in previously explored high-energy collisions---do not meet the divergence condition, an explicit example of a doubly divergent collision scenario has thus far gone unrealized.) 

In Fig.~\ref{fig:Collisions}, we show the CM energy of Type T and V collisions including IBCO$^+$ and ISCO$^+$ particles (orbiters and plungers). In particular, we showcase collision scenarios utilizing the exemplar particles in Tab.~\ref{tab:NearHorizonNearCriticalParams}, where a generic particle and an IBCO$^+$/ISCO$^+$ particle collide, or an IBCO$^+$ particle and ISCO$^+$ particle collide. Explicitly, we plot the normalized CM energy,
\begin{align}
    \label{eq:NCMEF}
    \mathcal{N}(E_{\cm})=\sqrt{\frac{E_{\cm}^2}{m_1 m_2}-\frac{m_1^2+m_2^2}{m_1m_2}},
\end{align}
which is independent of the particle masses, multiplied by the appropriate factor of $\kappa$ that regularizes the CM energy in the near-extremal limit $\kappa\to0$. Additionally we label the collision configurations appearing in prior work: JS Ref.~\cite{Jacobson2010}, GP Refs.~\cite{Grib2010,Grib2011a}, and HK Refs.~\cite{Harada2011a,Harada2011b}. 

Generic particles ($l_i\neq 0,\ q_i=0$) do satisfy the divergence condition, so the collision of a near-horizon circular orbit particle and a generic particle has divergent CM energy (see Fig.~\ref{fig:Collisions} panels, top and middle panels). But, the collision of two near-horizon circular orbit particles has finite CM energy (see Fig.~\ref{fig:Collisions}, bottom panels).
Next, we calculate the explicit form of CM energy for high-energy collisions including near-horizon circular orbit particles, in particular Type T collision with a IBCO$^+$/ISCO$^+$ orbiter and Type V collision at the horizon that include IBCO$^+$/ISCO$^+$ plungers (see Fig.~\ref{fig:Collisions}, dots along $\kappa=0$ axis in the top and middle panels).

\subsection{Type T collisions with circular orbiters}
\label{sec:TypeTCollisions}

The general formula for the CM energy of a singularly sourced divergent Type T collision with a near-horizon circular orbiter ($p_c=p_\circ$ and $q_2=q_\circ$) is given by
\begin{subequations}
\begin{align}
    \label{eq:TypeTCircOrbs}
    \frac{E_\mathrm{cm\near}^2}{m_1m_2}&= \dfrac{8l_1 \mu_\circ}{R_{\circ}\kappa^{p_c-q_1}}.
\end{align}
\end{subequations}
The strongest CM energy divergence occurs when the circular orbiter is in the near-NHEK band ($p_c=1$) and the ingoing particle is generic ($l_1\neq0,\ q_1=0$). In general, we find that the CM energy of a collision between a generic particle and a circular orbiter diverges according to the $p$-band in which the orbiter resides, 
\begin{align}
    E_\mathrm{cm\near}\propto\kappa^{-p_c/2} =\pa{1-a^2/M^2}^{-p_c/4},
\end{align}
explaining the CM energy divergence seen in previously studied collisions with IBCO$^+$ and ISCO$^+$ orbiters. When the orbiter is at the IBCO$^+$/ISCO$^+$ radius, the CM energy is
\begin{align}
    \label{eq:TypeTCircOrbs}
    \frac{E_\mathrm{cm\near}^2}{m_1m_2}=4l_1
    \begin{cases}
        \dfrac{\sqrt{2}}{\kappa}, & \Centerstack{$\text{IBCO}^+$ \text{orbiter}}\\[3ex]
        \dfrac{2^{2/3}}{\sqrt{3}\kappa^{2/3}}, & \Centerstack{$\text{ISCO}^+$ \text{orbiter}}
    \end{cases}.
\end{align}
The IBCO$^+$ case reproduces the finding of JS (Fig.~\ref{fig:Collisions}, top left panel), and the ISCO$^+$ case matches the findings of HK (Fig.~\ref{fig:Collisions}, middle left panel) \cite{Jacobson2010,Harada2011a,Harada2011b}. The appropriate values for particle 1 from Tab.~\ref{tab:NearHorizonNearCriticalParams} can be used to confirm~\footnote{\label{ft:comparison}For comparison to HK, our quantities $\cu{\mu_1,l_1}$ are equivalent to $\cu{e_2,1-l/e_2}$ of Refs.~\cite{Harada2011a,Harada2011b}.\\ For comparison to JS and GP, the near-extremal expansion in Refs.~\cite{Jacobson2010,Grib2010,Grib2011a} are performed in terms of $\epsilon=1-a/M\ll 1$. Thus, our small expansion parameter $\kappa\approx\sqrt{2\epsilon}$.}.

\subsection{Type V collisions with circular orbit plungers at the horizon}
\label{sec:TypeVCollisions}

The general formula for the CM energy of a singly sourced divergent Type V collision with a circular orbit plunger ($p_c\geq p_\circ$ and $q_2=2-p_\circ\geq1$) which occurs in the deepest part of the geometry $\pa{p_c=1}$ is given by
\begin{subequations}
\begin{align}
    \label{eq:SinglyDivergentTypeVPlunge}
    \frac{E_\mathrm{cm\near}^2}{m_1m_2}=&\frac{4 g -4\sqrt{g^2-l_1^2\pa{R^2-1}\pa{1+\mu_2^{2}+\eta_2}}}{\pa{1+\cos^2\theta}\pa{{R^2-1}}\kappa^{1-q_1}},\\
    g=&2l_1 R \mu_2 +l_2 \delta_{p_c,1}.
\end{align}
\end{subequations}
When particle 1 is generic ($l_1\neq 0,\ q_1=0$), we see the characteristic $E_\cm\propto \pa{1-a^2/M^2}^{-1/4}$ divergence.
Despite the factor of $\pa{R^2-1}$ in the denominator of \eqref{eq:SinglyDivergentTypeVPlunge}, as in the ISCO$^+$ and IBCO$^+$ plunger cases \eqref{eq:SinglyDivergentIBCO_ISCO}, the CM energy of collisions is well defined at the horizon. Taking the near-horizon limit $R-1\ll 1$, \eqref{eq:SinglyDivergentTypeVPlunge} becomes
\begin{align}
    \frac{E_\mathrm{cm\near}^2}{m_1m_2}=&\frac{l_1\pa{1+\mu_2^2+\eta_2}}{\pa{\mu_2+l_2\delta_{q_2,1}}\pa{1+\cos^2\theta}\kappa^{1-q_1}}.
\end{align}

If particle 2 is an ISCO$^+$/IBCO$^+$ plunger, the CM energy reduces to
\begin{align}
    \label{eq:SinglyDivergentIBCO_ISCO}
    \frac{E_\mathrm{cm\near}^2}{m_1 m_2}=&\dfrac{4 l_1}{\pa{R+1}\kappa^{1-q_1}}
    \begin{cases}
        2+\sqrt{2}, & \Centerstack{$\text{IBCO}^+$ \text{plunger}}\\[3ex]
        \dfrac{2}{\sqrt{3}}, &
        \Centerstack{$\text{ISCO}^+$ \text{plunger}}
    \end{cases}.
\end{align}
When particle 1 is generic ($q_1=0$) and the collision occurs at the horizon ($R=1$), 
\begin{align}
    \label{eq:TypeVCircOrbs}
    \frac{E_\mathrm{cm\near}^2}{m_1 m_2}=&\dfrac{2 l_1}{\kappa}
    \begin{cases}
        2+\sqrt{2}&
        \Centerstack{$\text{IBCO}^+$ \text{plunger}}\\[3ex]
        \dfrac{2}{\sqrt{3}}, &
        \Centerstack{$\text{ISCO}^+$ \text{plunger}}
    \end{cases}.
\end{align}
The IBCO$^+$ case reproduces the finding of JS and GP (Fig.~\ref{fig:Collisions}, top right panel), and the ISCO$^+$ case matches the findings of HK (Fig.~\ref{fig:Collisions}, middle right panel)
\cite{Jacobson2010,Grib2010,Grib2011a,Harada2011a}. The appropriate values for particle 1 from Tab.~\ref{tab:NearHorizonNearCriticalParams} can be used to confirm~${}^{\ref{ft:comparison}}$.

\subsection{Type V collisions with circular orbit plungers in the ($\bf{0<p<1}$) bands}
\label{sec:TypeVCollisions2}

The general formula for the CM energy of a singly sourced divergent Type V collision with a circular orbit plunger ($p_c\geq p_\circ$ and $q_2=2-p_\circ>1$) which occurs in a $0<p<1$ band of the geometry is given by
\begin{align}
    \label{eq:SinglyDivergentTypeVPlunge2}
    \frac{E_\mathrm{cm\near}^2}{m_1m_2}=&\frac{4 \pa{2l_1\mu_2-\sqrt{l_1^2}\sqrt{3\mu_2^2-1-\eta_2}}}{\pa{1+\cos^2\theta}R\kappa^{p_c-q_1}}.
\end{align}
When particle 2 is an ISCO$^+$ plunger, the CM energy further simplifies to
\begin{align}
    \frac{E_\mathrm{cm\near}^2}{m_1m_2}=&\frac{8 l_1}{\sqrt{3}R\kappa^{p_c-q_1}}.
\end{align}

Additionally we note that Eq.~\eqref{eq:SinglyDivergentTypeVPlunge2} is also the form of the CM energy for singularly divergent collisions in the $0<p<1$ bands that include precisely critical particles or near-critical particles with $q_i>p_c$ such as a particle that will asymptotically spiral onto unstable circular orbits (i.e., particles with $\mu=\mu_\circ$ and $L=L_\circ(r_\circ)$ at radii $r>r_\circ$). (A detailed description of high-energy collision with precisely critical particle is given in App.~\ref{app:PreciselyCritical}.)

In Ref.~\cite{Banados2009}, BSW numerically showed the CM energy divergence of the Type V collision of a IBCO$^-$ plunger and an equatorially bound ($\eta_2=0$) marginally bound ($\mu_2=0$) precisely critical particle on the extreme BH background. Evaluating Eq.~\eqref{eq:SinglyDivergentTypeVPlunge2} for particle 1 as an IBCO$^-$ plunger ($l_1=-1/\sqrt{2}$ and $q_1=0$) and particle 2 as an equatorially bound, marginally bound precisely critical particle ($\eta_2=0$ and $\mu_2=1$) or a particle which will spiral onto the IBCO$^-$ if undisturbed, we find
\begin{align}
    \frac{E_\mathrm{cm\near}^2}{m_1m_2}=&\frac{4\pa{2-\sqrt{2}} l_1}{R\kappa^{p_c-q_1}}.
\end{align}
This CM energy matches the leading order behavior of the BSW collision in the near-horizon limit $\epsilon=(r/M-1)\to0$ under the mapping of $R\kappa^{p_c-q_1}\mapsto \epsilon$ (see App.~\ref{app:EKCMEnergy}). This correspondence of the BSW exemplar Type V collision of the on-horizon collision of the extremal BH and the Type V collision in the $0<p<1$ bands of the near-extremal BH is to be expected as the near-horizon region of the extremal BH and be $0<p<1$ bands of the near-extremal BH are both described by the NHEK metric (see App.~\ref{app:NHEK}).

\section{Naturalness and observability of high-energy collisions} 
\label{sec:Discussion}

We have performed a theoretical investigation to determine the conditions under which divergent CM energy collisions occur around near-extremal BHs, and have derived the explicit form of the CM energy for these high-energy collisions.
While, experimental evidence suggest several astrophysical BHs are rotating quite rapidly \cite{Reynolds2019,Draghis2023}, astrophysical BHs are believed to be sub-extremal due to physical processes like accretion and back-reaction \cite{Thorne1974, Berti2009}.
Moreover, the CM energy grows slowly with $\kappa$, so the enhancement of the CM energy for rapidly rotating BHs is modest unless a very high spin cutoff is imposed. 

Certainly generic particles falling toward a BH horizon are easily realized. To realize the critical particles, we can consider an accretion disk around a BH. In the case of radiatively efficient accretion flow, the accretion disk may settle into a geometrically thin, equatorial configuration in which particles with radii at or above the ISCO$^+$ radius ($r\geq r_{\mathrm{isco}}^+$) travel on (quasi-)stable circular orbits, while particles interior to the ISCO$^+$ radius ($r<r_{\mathrm{isco}}^+$) plunge toward the BH while maintaining the ISCO$^+$ orbiter energy and angular momentum ($\mu=\mu_{\mathrm{isco}}^+$ and $L=L_{\mathrm{isco}}^+$) \cite{Novikov1973,Cunningham1975}. This disk model 
underlies X-ray spectroscopic methods used to constrain the spins of supermassive BH at the heart of galaxies and stellar-mass BHs in X-ray binaries \cite{Zhang2024,Brenneman2006,Fabian2000,Reynolds2020}. In this disk model, when the BH is near-extremal, the near-horizon circular orbiters can participate in Type T collisions, and plunging particles of the interior disk can participate in Type V collisions. 

\begin{figure}
    \centering
    \includegraphics[width=.5\columnwidth]{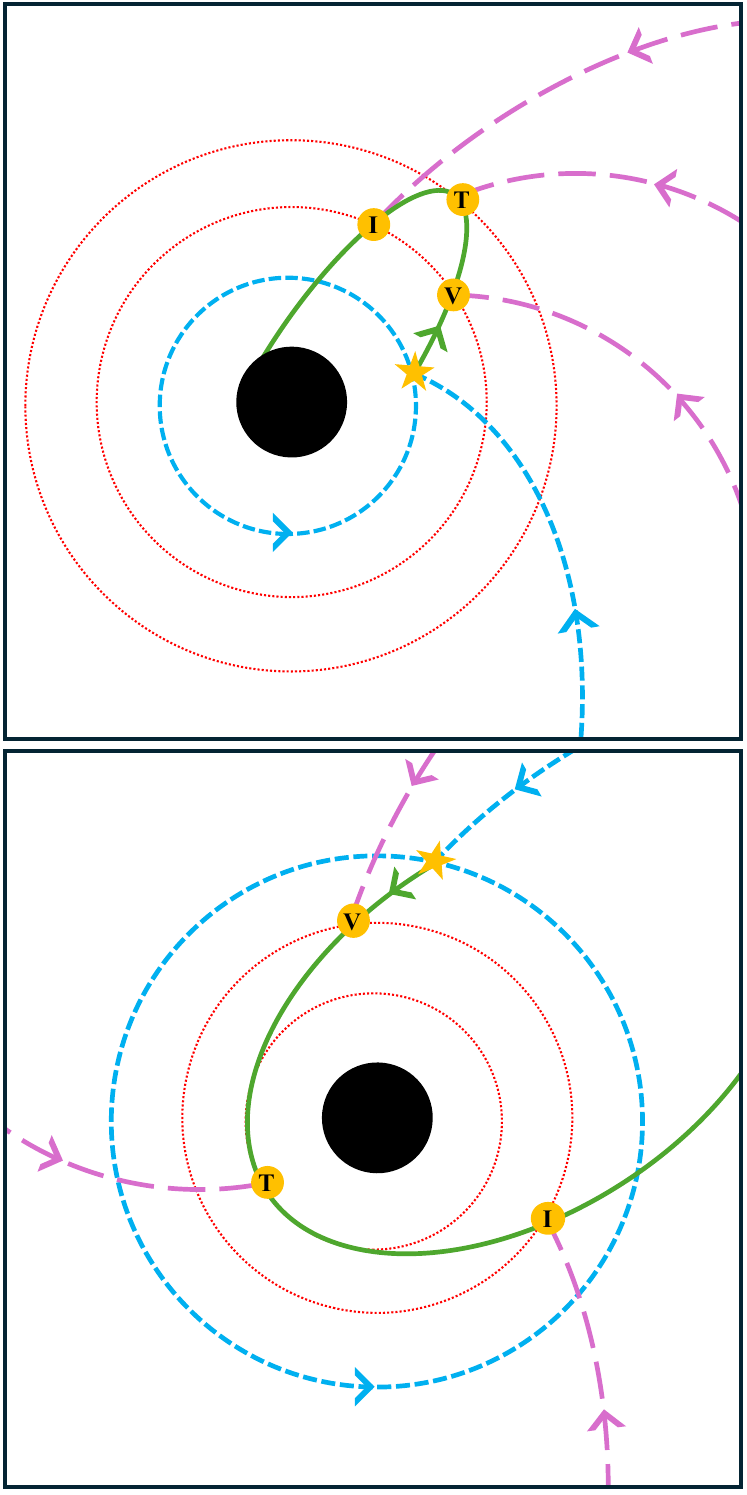}
    \caption{Schematic depictions of multi-scattering scenarios near BH. An initial Type T collision occurs between a circular orbit particle and an infalling particle (trajectories show as short-dashed blue curves with collision location marked by yellow stars). A daughter particle of the initial collision is emitted with outward (left panel) or inward (right panel) radial momentum component, traveling along a trapped or deflecting geodesic (solid green curves). The daughter particle may participate in a secondary collision with another infalling particle. The secondary collision Type depends on the location at which the daughter particle is intercepted. We show the trajectories of three infalling particles (long-dashed purple curves) which can participate in the secondary collision with the daughter particle. The respective secondary collision radii are marked as dotted red curves with the collision locations marked by yellow circles that are labeled by collision Type. The BH is shown as a black circle.}
    \label{fig:DaughterCollision}
\end{figure}

Thus far, we have not discussed Type I collisions, which require a particle to have an outward radial momentum component at the time of collision. 
The (near-)NHEK geometries admit particle motion with both outward radial momentum component. 
In particular, timelike geodesics on the in the (near-)NHEK geometries can be described in five broad classes of radial behavior: 1) plunging geodesics with no radial turning point which connect infinity to the horizon, 2) outward geodesics with no radial turning point which connecting the horizon to infinity, 3) spherical orbits that maintain fixed radius 3) trapped geodesics with one radial turning point which connect the horizon to the horizon, and 5) defecting (osculating) geodesics which connect infinity to infinity \cite{Hadar2014,Hadar2017,Kapec2020,Compere2020}. 
Hence particles on outward, trapped, and deflecting orbits may
participate in in Type I collisions with infalling generic particles. 
While we have established that Type I collision in the near the horizon can exist in principle, one needs an astrophysical set up in which to realize near-horizon particles with outward radial motion component. Once such a set up is identified precise divergences would depend on being able to calculate the rate at which the constituent particles scale to criticality as the near-extremal limit of orbits in the Kerr geometry, $l\kappa^q\approx\mu-L/(2M)$ for $\kappa\ll1$.

Multi-scattering scenarios may seed additional near-horizon collisions, including Type I collisions. It has been shown that the vast majority of emission from the high-energy collision of a generic particle and a circular orbiter is tuned to criticality at the same rate as the orbiter (see App. B of Ref.~\cite{Gates2023}). Therefore, a daughter particle of a circular orbiter collision could certainly collide with, say, another generic infalling particle in a singular divergent collision. If we consider an initial Type T collision of an infalling generic particle and a near-horizon circular orbit particle in a geometrically thin accretion disk, a daughter particle born of the initial collision can participate in a secondary collision with another infalling generic particle.
Daughter particles on plunging orbits ($p_r<0$ always) can participate in Type V collisions. Daughter particles on spherical orbits ($p_r=0$ always) can only participate in Type T collisions~\footnote{These may be less common as only a co-dimension one set of emission directions in the collision CM frame correspond to $p_r=0$.}. Daughter particles on outward orbits ($p_r>0$ always) only can participate in Type I collisions. Lastly, daughter particles on a trapped or deflecting geodesic can participate in any collision Type depending on where along its trajectory the infalling generic particle impinges upon it (see Fig.~\ref{fig:DaughterCollision}).
Determining the form of (and divergence conditions for) the secondary collision CM energy requires specifying the energy and angular momentum of the daughter particle as a function of the initial collision parameters (i.e., collision radius, specific energies, and angular momenta), which requires specifying the scattering process that produces the daughter particle, and is an investigation for future inquiry.

Finally, high-energy near-horizon collisions can be observable. Despite the fact that the collisions occur in the near-horizon region, daughter particles of these high-energy collisions can escape to infinity where they can be detected. The near-horizon scattering processes examined to date have resulted in daughter particles with observed energy as high as the rest mass of the parent particles \cite{Piran1975,Piran1977,Piran1977a, Jacobson2010,Bejger2012,Leiderschneider2016, Liberati2022, Gates2023}.

\section*{acknowledgments}
DG is grateful to Shahar Hadar for insightful discussions and detailed comments on this work. 
D.G. acknowledges support from the Harvard Postdoctoral Fellowship for Future Faculty Leaders. D.G. acknowledges financial support from the National Science Foundation (AST-2307887). This publication is funded in part by the Gordon and Betty Moore Foundation (Grant \#8273.01). It was also made possible through the support of a grant from the John Templeton Foundation (Grant \#62286). The opinions expressed in this publication are those of the author and do not necessarily reflect the views of these Foundations. While at Princeton, D.G. was supported in this work by a Princeton Gravity Initiative postdoctoral fellowship and by a Princeton Future Faculty in the Physical Sciences fellowship.

\appendix

\section{High-energy collisions on the extreme Kerr background}
\label{app:EKCMEnergy}

The extreme Kerr geometry is given by \eqref{eq:Kerr} with $a=M$. The horizon is at radius $r_\mathrm{H}=M$ and has angular velocity $\Omega_\mathrm{H}=1/(2M)$. Setting $a=M$ and taking the near-horizon limit $r\to M$ in the radial CM energy term \eqref{eq:Ecm}, we find
\begin{align}
    \label{eq:ECMchiNearHorizonEK}
    \chi=
    \begin{cases}
        \dfrac{T_2\hat L_1}{2\hat L_2}+\dfrac{T_1 \hat L_2}{2\hat L_1}+\O{r-M}, & s_{r}=1\\[3ex]
        \dfrac{\pa{1-s_{r}}M^2\hat L_1\hat L_2}{\pa{r-M}^2}+\O{\pa{r-M}^{-1}}, & s_{r}\leq0
    \end{cases},
\end{align}
where $\hat L_i=L_i-2M\mu_i$. 
Note that Types T and I collisions diverge faster than in the sub-extremal case $\chi$ \eqref{eq:ECMchiNearHorizon}, while the Type V collisions have the same divergence. As in the sub-extremal case, tuning the particles to criticality may change the divergence of the CM energy; however, this formula is only guaranteed to apply when both particles scale to criticality slower than the collision radius approaches the horizon. To carefully perform the near-horizon limit with particles whose angular momentum may scale to criticality at various rates relative to the collision radius, we develop a multi-scaling limit. 

To resolve $\chi$ \eqref{eq:ECMchi} to leading order, we must resolve $r-M$ and $L-2M\mu$ to leading order; thus, we introduce the ansatz,
\begin{subequations}
\label{eq:ScalingLimitEKAnsatz}
\begin{alignat}{2}
    a &=M,&       & \\
    \label{eq:RScalingEK}
    r   &= M(1+\epsilon), &\quad  & \epsilon\ll1, \\
    \label{eq:LScalingEK}
    L_i &= 2M\pa{\mu_i-\ell_i \epsilon^{n_i}}, &\quad  & 0\leq n_i.
\end{alignat}
\end{subequations}
Eq.~\eqref{eq:LScalingEK} allows us to describe three types of particles: generic particles ($\ell_i\neq0$, $n_i=0$), near-critical particles ($\ell_i\neq 0$, $n_i>0$), and precisely critical particles ($\ell_i=0$). Using this ansatz, we can define the extremal, near-critical, near-horizon multi-scaling limit of a quantity $f$ as
\begin{align}
    \label{eq:ScalingLimitEK}
    f(a,r,L_1,L_2)\Big\vert_{\text{Eq.\eqref{eq:ScalingLimitEKAnsatz}}} \overset{\epsilon \to 0}{\sim} \ f_{\ext}(\epsilon).
\end{align}

The (possible) divergence of $\chi$ \eqref{eq:ECMchi} can be diagnosed by examining $\tilde P$ and $T$.
In the multi-scaling limit \eqref{eq:ScalingLimitEK}, 
\begin{align}
    \tilde P_{i\ext}&=2M
    \begin{cases}
        \dfrac{\ell_i}{\epsilon^{1-n_i}}, &0\leq n_i<1, \ell_i\neq0\\[2ex]
        \mu_i+\ell_i, & n_i=1, \ell_i\neq0\\[1ex]
        \mu_i, & (n_i>1,\ell_i\neq0) \text{ or }(\ell_i=0)
    \end{cases},
    \label{eq:Pext}
\end{align}
and
\begin{align}
    T_{i\ext}&=
        \tilde{T}_{i\ext}+4M^2 \ell_i(\ell_i-\mu_i)\delta_{n_i,0}\\
    \label{eq:Text}
    \tilde{T}_{i\ext}&=M^2\pa{1+\eta_i+\mu_i^{2}},
\end{align}
where $\eta_i=Q/M^2$. Since $\tilde{P}_{i\ext}$ is non-vanishing and $T_{i\ext}$ is finite, $R_i\propto\tilde P_i^2-T_i\geq0$ imposes no condition on $n_i$.

Divergent CM energy requires at least one particle satisfies
$$\text{divergence condition:}\ \ell_i\neq0,\ n_i<1 .$$ 
For singly sourced divergent collisions of all Types, wherein particle 1 satisfies the divergence condition,
\begin{align}
    \label{eq:SinglySourcedExt}
    \chi_{\ext}&=\tilde P_{1\ext}\tilde P_{2\ext}-s_{r}\sqrt{\tilde P_{1\ext}^2}\sqrt{\tilde P_{2\ext}^2-\tilde T_{2\ext}}\propto\frac{1}{\epsilon^{1-n_1}}.
\end{align}
For doubly sourced divergent collisions of Types T and I, wherein $n_1\leq n_2$,
\begin{align}
    \label{eq:DoublySourcedTypeIandTExt}
    \chi_{\ext}=(1-s_{r})\tilde P_{1\ext}\tilde P_{2\ext}\propto\frac{1}{\epsilon^{2-n_2-n_1}}.
\end{align}
For doubly sourced divergent collisions of Type V, wherein $n_1<n_2$,
\begin{align}
    \label{eq:DoublySourcedTypeVExt}
    \chi_{\ext}=\frac{\tilde T_{2\ext}{\tilde P_{1\ext}} }{2{\tilde P_{2\ext}}}\propto\frac{1}{\epsilon^{n_2-n_1}}.
\end{align}
The doubly divergent collision formulae Eqs.~\eqref{eq:DoublySourcedTypeIandTExt} and \eqref{eq:DoublySourcedTypeVExt} match Eq.~\eqref{eq:ECMchiNearHorizonEK} substituting $L_i$ of the form \eqref{eq:LScalingEK}. The singly divergent collision formula Eq.~\eqref{eq:SinglySourcedExt} cannot be deduced from Eq.~\eqref{eq:ECMchiNearHorizonEK}. 

In Ref.~\cite{Banados2009}, BSW analytically presented the formula for the near-horizon \emph{doubly divergent} Type V collisions with equal mass $\mu=1$ particles given on the equatorial plane, which corresponds to $\chi$ \eqref{eq:ECMchiNearHorizonEK} with $s_r=1$ herein. This equation allowed them to deduce the existence of divergent the Type V collisions and identify the critical momentum $L=2M\mu$. BSW also demonstrated numerically that the CM energy of a Type V collision between a generic particle and a precisely critical particle diverges as the collision approaches the horizon, although they did not derive its explicit form. We will now derive the the analytic form of the \emph{singularly divergent} collision studied by BSW.

The exemplar collision used by BSW featured a particle plunging from the retrograde IBCO (i.e., $(\mu_1,L_1/M)=(1,-2-2\sqrt{2})$) corresponding to $(\ell_1/\mu_1,n_1)=(2+\sqrt{2},0)$) and a marginally bound precisely critical particle (i.e., $(\mu_2,L_2/M)=(1,2)$ corresponding to $\ell_i=0$)~\footnote{Ostensibly, BSW limited their analysis to the ``direct collision scenario'' in which the colliding particles fall in from infinity with $\mu=1$. For the extremal BH, particles with $L\in[-2(1+\sqrt{2})M,2M]$ can reach the horizon. BSW examined the collision of particles with the maximal and minimal angular momenta with which particles can reach the horizon. But the $L=-2(1+\sqrt{2})M$ particle falling in from infinity asymptotically spirals onto the IBCO$^-$ radius, whereas the horizon reaching particle asymptotically spirals off the IBCO$^-$ radius and plunges into the BH \cite{Compere2021}.}.
The collision of a generic particle ($\ell_1\neq0,\ n_1=0$) with either a precisely critical particle ($\ell_2=0$) or a near-critical particle with $n_2>1$ has near-horizon CM energy
\begin{align}
    \label{eq:TypeVPCExt1}
    \frac{E_{\cm\ext}^2}{m_1m_2}=\frac{4\pa{2\ell_1\mu_2-\sqrt{\ell_1^2}\sqrt{3\mu_2^2-(1+\eta_2)}}}{\pa{1+\cos^2\theta}\epsilon}.
\end{align}
Furthermore, when the critical particle is equatorially bound ($\eta_2=0$) and marginally bound ($\mu_2=1$),
\begin{align}
    \label{eq:TypeVPCExt}
    \frac{E_{\cm\ext}^2}{m_1m_2}=\frac{4\ell_1\pa{2-\sqrt{2}}}{\epsilon}.
\end{align}
Finally, taking the generic particle to be a IBCO$^-$ plunger, we recover the near-horizon CM energy of the BSW exemplar collision. We show examples of the divergent CM energy Type V collision of a precisely critical particle and a generic particle in Fig.~\ref{fig:Collisions} (top panel).

\begin{figure}
    \centering
    \includegraphics[width=.5\linewidth]{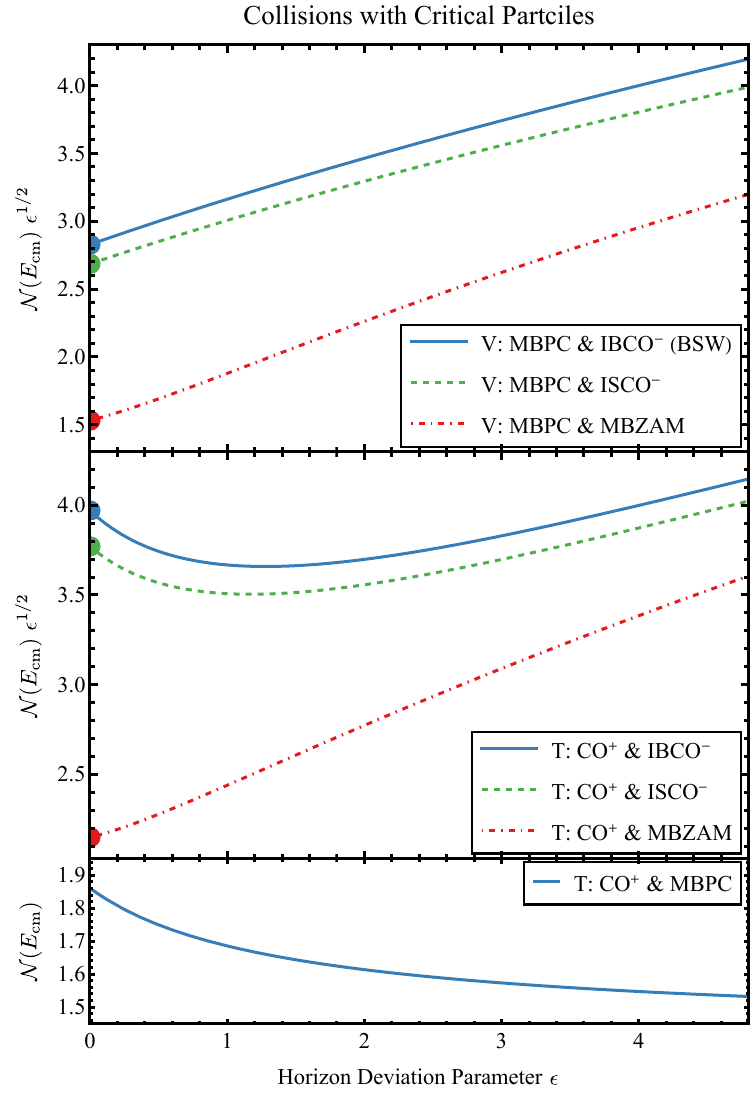}
    \caption{CM energies of collision including precisely critical particles and prograde circular orbiters (near-critical particle). We plot the a normalized CM energy $\mathcal{N}(E_{\cm})$ \eqref{eq:NCMEF} as a function of horizon deviation parameter $\epsilon=1-r/M$. The normalized CM energy has been multiplied by the appropriate factors of $\epsilon$ which regularize the CM energy in the near-horizon limit $\epsilon\to0$.\\[1ex]
    {\bf Top panel:} Type V collisions at the horizon of equatorial bound, marginally bound precisely critical particles (labeled ``MBPC'') and generic particles: retrograde IBCO plunger (``IBCO$^-$''), ISCO plunger (``ISCO$^-$''), and marginally bound zero angular momentum particle (``MBZAM'').\\ 
    {\bf Middle panel:} Type T collisions of prograde circular orbiter (``CO$^+$'') and generic particles.\\
    {\bf Bottom panel:} Type T collision of a prograde circular orbiter and an equatorially bound, marginally bound precisely critical particle.\\[1ex]
    In the near-horizon limit, the CM energy of the Type V collision of a precisely critical particle and a generic particle and the CM energy of the Type T collision of a circular orbiter and a generic particle diverge ($E_{\cm}\sim\mathcal{O} (\epsilon^{-1/2})$ for $\epsilon\to0$). 
    The analytic formulae for the high-energy CM energy $E_{\cm \ext}/\sqrt{m_1m_2}$ for the Type V and T collision are given by Eqs.~\eqref{eq:TypeVPCExt} and \eqref{eq:TypeTCircOrbsExt}, respectively, and are illustrated as dots along the $\epsilon=0$ axis in the top and middle panels. The CM energy of a collision of the a prograde circular orbiter and a precisely critical particle is finite in the near-horizon limit.
    }
    \label{fig:CollisionsExtreme}
\end{figure}

The extremal Kerr BH geometry also admits prograde stable equatorial circular orbits at all radii outside the horizon, $r>r_{H}$ with specific and angular momentum given by Eq.~\eqref{eq:CircularOrbiter} with $a=M$. As such, we can examine Type T collisions with circular orbiters. The circular orbiter has near-horizon limit $\pa{\mu_\circ^+,L_\circ^+/M}=(1/\sqrt{3},2/\sqrt{3})$. 
By \eqref{eq:LScalingEK}, the near-horizon limit $\ell_\circ\epsilon^{n_\circ}\approx\mu_\circ^+-L_\circ^+/(2M)$ implies $(\ell_\circ,n_\circ)=(-1/4,2)$; thus, the near-horizon circular orbiter is a near-critical particle.
As $n_\circ>1$, the collision of a generic particle ($\ell_1\neq0,\ n_1=0$) and the near-horizon circular orbit has a singly sourced divergent CM energy 
\begin{align}
    \label{eq:TypeTCircOrbsExt}
    \frac{E_{\cm\ext}^2}{m_1m_2}=\frac{8\ell_1}{\sqrt{3}\epsilon},
\end{align}
(see Fig.~\ref{fig:CollisionsExtreme}, middle panel).
Further, the collision of a precisely critical particle and the near-horizon orbiter is finite (see Fig.~\ref{fig:CollisionsExtreme}, top panel).
Thus, we find the same behavior as the near-extremal case Sec.~\ref{sec:TypeTCollisions}, wherein near-horizon Type T collisions are divergent (finite) when the incoming particle is generic (precisely critical) as well, e.g., Fig.~\ref{fig:Collisions} left top and middle panels (left bottom panel). 

On the extremal background, circular orbits and ingoing generic and precisely critical particles that start far from the BH and fall inward are easy to realize astrophysically. Therefore, while extremal BHs are not believed to exist in our universe~\cite{Thorne1974,Berti2009}, if they were to exist, we might reasonably expect high-energy collisions to take place around them. 

Finally, we note the high-energy collisions near the extremal BH can be mapped to high-energy collisions in the $0<p<1$ band of the near-extremal BH (see Sec.~\ref{sec:TypeVCollisions2}). Comparing Eqs.~\eqref{eq:RScalingEK}-\eqref{eq:LScalingEK} to Eqs.~\eqref{eq:RScaling}-\eqref{eq:LScaling}, we find the transformation
\begin{align}
    \label{eq:MapToEK}
    \epsilon\mapsto R\kappa^{p_c},\quad \ell_i\mapsto \frac{l_i}{R^n\kappa^{np_c-q_i}}.
\end{align}
Additionally, comparing Eqs.~\eqref{eq:Pext}-\eqref{eq:Text} to Eqs.~\eqref{eq:TildesPnear}-\eqref{eq:Tnear} with $0<p_c<1$,
\begin{align}
    \text{Eq.~\eqref{eq:MapToEK}: } \pa{\tilde P_{i\ext},T_{i\ext}}\mapsto \pa{\tilde P_{i\near},T_{i\near} },
\end{align}
if we impose the additional condition that $n_i=q_i/p_c$. Hence, we can map near-horizon collisions on the extremal to near-horizon collision in the $0<p<1$ band of the near-extremal background using the transformation
\begin{align}
    \label{eq:MapToEK2}
    \epsilon\mapsto R\kappa^{p_c},\quad \ell_i\mapsto \frac{l_i}{R^{q_i/p_c}}, \quad n_i=\frac{q_i}{p_c}.
\end{align}
Indeed, applying this transformation to Eq.~\eqref{eq:TypeVPCExt1} results in Eq.~\eqref{eq:SinglyDivergentTypeVPlunge2}, confirming the mapping of the Type V collision of a generic particle and precisely critical particle/near-critical particle with $n_i=q_i/p_c>1$ on the two backgrounds. Likewise, applying this transformation to Eq.~\eqref{eq:TypeTCircOrbsExt} yields Eq.~\eqref{eq:TypeTCircOrbs} with $p_\circ<1$, confirming the mapping of the Type T collision of a generic particle and a circular orbiter on the two backgrounds.

\section{High-energy collision on the sub-extremal Kerr background}
\label{app:SubEKCMEnergy}

The Kerr geometry is given by Eq.~\eqref{eq:Kerr} with the horizon at radius \eqref{eq:Horizon}. In the CM energy \eqref{eq:Ecm} only the radially dependent term \eqref{eq:ECMchi} may be divergent in the near-horizon limit. Eq.~\eqref{eq:ECMchiNearHorizon} suggests that tuning a particle to criticality will affect the CM energy divergence of collisions, but strictly speaking is only guaranteed to be applicable to collision involving a particle with angular momenta approach criticality slower than the collision radius approaches the horizon. To carefully consider near-critical particles, i.e., particles whose angular momenta approach criticality as the collision approaches the horizon, we must perform a multi-scaling limit.

To resolve $\chi$ \eqref{eq:ECMchi} to leading order, we must resolve $r-r_\mathrm{H}$ and $L-\Omega_\mathrm{H}^{-1}\mu$ to leading order. Thus, we adopt the ansatz
\begin{subequations}
\label{eq:ScalingLimitSubAnsatz}
\begin{alignat}{2}
    a &< M, && \\
    \label{eq:RScalingSubEK}
    r &=r_\mathrm{H} + M\epsilon, &\quad  & 0<\epsilon\ll1,\\ 
    \label{eq:LScalingSubEK}
    L_i &= \Omega_{\mathrm{H}}^{-1}\pa{\mu_i-\ell_i \epsilon^{n_i}}, &\quad  & 0\leq n_i.
\end{alignat}
\end{subequations}
Eq.~\eqref{eq:LScalingSubEK} allows us to describe three types of particles: generic particles ($\ell_i\neq0$, $n_i=0$), near-critical particles ($\ell_i\neq 0$, $n_i>0$), and precisely critical particles ($\ell_i=0$).
Using this ansatz, we can define the near-extremal, sub-critical, near-horizon multi-scaling limit of a quantity $f$ as
\begin{align}
    \label{eq:ScalingLimitSub}
    f(a,r,L_1,L_2)\Big\vert_{\text{Eq.\eqref{eq:ScalingLimitSubAnsatz}}} \overset{\epsilon \to 0}{\sim} \ f_{\sub}(\epsilon).
\end{align}

The (possible) divergence of $\chi$ \eqref{eq:ECMchi}, can be diagnosed by examining $\tilde P$ and $T$.
To leading order in the multi-scaling limit, 
\begin{subequations}
\label{eq:PtildeSub}
\begin{align}
    \tilde P_{i\sub}&=\frac{r_{\mathrm{H}}\sqrt{2 M \epsilon}}{\pa{M^2-a^2}^{1/4}} \mathcal{A}_\ell,\\
    \mathcal{A}_\ell&=
    \begin{cases}
        \dfrac{\ell_i}{\epsilon^{1-n_i}}, &0\leq n_i<1,\ell_i\neq0\\[2ex]
        \mu_i+\ell_i, & n_i=1,\ell_i\neq0\\[1ex]
        \mu_i, & (n_i>1,\ell_i\neq0) \text{ or }(\ell_i=0)
    \end{cases},
\end{align}
\end{subequations}
and
\begin{align}
    \label{eq:TSub}
    T_{i\sub}&=\tilde{T}_{i\sub} + \dfrac{r_\mathrm{H}^2}{a^2} 4M\ell_i (M\ell_i-r_\mathrm{H}\mu_i)\delta_{n_i,0},\\
    \tilde{T}_{i\sub}&=Q_i+r_{\mathrm{H}}^2 \pa{1+\frac{\mu_i^2 r_\mathrm{H}^2}{a^2}}.
\end{align}

Particle motion is only allowed when $R_i\propto\tilde P_i^2-T_i\geq0$. $T_{i\sub}$ is always finite and non-vanishing, so we must additionally restrict $n_i\leq1/2$ such that $\tilde P_{i\sub}$ is non-vanishing.

To produce divergent CM energy, $\tilde{P}_i$ of at least one particle must satisfy 
$$\text{divergence condition:}\ \ell_i\neq0,\ n_i<1/2.$$ 
For singly sourced divergent collisions of all Types, wherein particle 1 satisfies the divergence condition,
$\chi$~\eqref{eq:ECMchi} to leading order is
\begin{align}
    \label{eq:SinglySourcedSub}
    \chi_{\sub}&=\tilde P_{1\sub}\tilde P_{2\sub}-s_{r}\sqrt{\tilde P_{1\sub}^2}\sqrt{\tilde P_{2\sub}^2-\tilde T_{2\sub}}\propto\frac{1}{\epsilon^{1/2-n_1}}.
\end{align}
For doubly sourced divergent collisions of Types T and I, wherein $n_1\leq n_2<1/2$,
\begin{align}
    \label{eq:DoublySourcedTypeIandTSub}
    \chi_{\sub}=(1-s_{r})\tilde P_{1\sub}\tilde P_{2\sub}\propto\frac{1}{\epsilon^{1-n_2-n_1}},
\end{align}
For doubly sourced divergent collisions of Type V, wherein $n_1<n_2$,
\begin{align}
    \label{eq:DoublySourcedTypeVSub}
    \chi_{\sub}=\frac{\tilde T_{1\sub}{\tilde P_{1\sub}} }{2{\tilde P_{2\sub}}}\propto\frac{1}{\epsilon^{n_2-n_1}},
\end{align}
This is the expected form of divergent CM energy one could deduce from \eqref{eq:ECMchiNearHorizon} as the particles scale to criticality slower than the collision radius scales to the horizon. (Note also, in contrast to the the extremal and near-extremal BH, the sub-extremal BH does not have have near-horizon circular orbiters \eqref{eq:CircularOrbiter} that can participate in high-energy Type T collisions.)

In Ref.~\cite{Grib2010}, GP showed that non-extremal BHs can also exhibit high-energy collision if multiple scattering is allowed, i.e., the colliding particles are not required to fall in from infinity, and presented the analytic formula for doubly divergent Type V collisions for equal mass $\mu=1$ particles, corresponding to Eq.~\eqref{eq:DoublySourcedTypeVSub}.
GP suggest that prior collisions with accretion disk particle or particle decay near the horizon may source a particle with the angular momentum tuning needed for a high-energy collision. Further studies are needed to determine the astrophysical feasibility of these mechanisms for seeding near-critical particles near a sub-extremal BH. 

Let us note here we have determined the conditions for, and CM energy of, high-energy collisions near a sub-extremal BH using an angular momentum scaling formalism in which we keep fixed the particle specific energies and scale their angular momenta to criticality at varying rates (just as we have for the extremal and near-extremal BHs). However, this formalism cannot be used for the Schwarzschild BH as Eq.~\eqref{eq:CriticalAngularMomentum} is ill-defined when $a=0$. Instead, to resolve the high-energy collision near the Schwarzschild BH, we utilize a specific energy scaling formalism.

\subsection{specific energy scaling criticality formalism}
\label{app:SchwarzschildCMEnergy}

The radially dependent term of the CM energy \eqref{eq:Kerr} can be written as
\begin{align}
    \label{eq:ECMchiNearHorizonSchwarz}
    \chi=
    \begin{cases}
        \dfrac{T_2\hat \mu_1}{2\hat \mu_2}+\dfrac{T_1 \hat \mu_2}{2\hat \mu_1}+\O{r-r_{\mathrm{H}}}, & s_{r}=1\\[3ex]
        \dfrac{\pa{1-s_{r}}2M^2 r_{\mathrm{H}}^2\hat \mu_1\hat \mu_2}{\sqrt{M^2-a^2}\pa{r-r_\mathrm{H}}}+\O{1}, & s_{r}\leq0
    \end{cases},
\end{align}
where $\hat \mu_i\equiv \mu_i-\Omega_\mathrm{H}L_i$. In this form, we see that criticality of a particle can be described by specific energy
\begin{align}
    \label{eq:CriticalSpecificEnergy}
    \mu_*\equiv\Omega_{\mathrm{H}}L_i,
\end{align}
which is well defined for all BH spins $0\leq a<M$.
Once again, however, the form of $\chi$ above is only guaranteed to apply when $\mu_i$ scale to criticality slower than the collision radius scales to the horizon.

To resolve $\chi$ \eqref{eq:ECMchi} to leading order, we must resolve $r-r_\mathrm{H}$ and $\mu-\Omega_\mathrm{H}L$ to leading order. Thus, we adopt the ansatz
\begin{subequations}
\label{eq:ScalingLimitSchwarzAnsatz}
\begin{alignat}{2}
    a &< M,   &       & \\[0.3em]
    \label{eq:RScalingScwarz}
    r &= r_\mathrm{H}+M\epsilon, &\quad  & 0<\epsilon\ll1, \\
    \label{eq:LScalingSchwarz}
    \mu_i &= \Omega_{\mathrm{H}}L_i + \frac{M e_i}{2r_\mathrm{H}} \epsilon^{n_i}, &\quad  & 0\leq n_i.
\end{alignat}
\end{subequations}
Eq.~\eqref{eq:LScalingSchwarz} allows us to describe three types of particles: generic particles ($e_i\neq0$, $n_i=0$), near-critical particles ($e_i\neq 0$, $n_i>0$), and precisely critical particles ($e_i=0$).
Using this ansatz, we can define the near-extremal, sub-critical, near-horizon multi-scaling limit of a quantity $f$ as
\begin{align}
    \label{eq:ScalingLimitSchwarz}
    f(a,r,L_1,L_2)\Big\vert_{\text{Eq.\eqref{eq:ScalingLimitSchwarzAnsatz}}} \overset{\epsilon \to 0}{\sim} \ f_{\sub}(\epsilon).
\end{align}

To leading order in this multi-scaling limit, \eqref{eq:ScalingLimitSchwarz}
\begin{subequations}
\label{eq:PtildeSchwarz}
\begin{align}
    \tilde P_{i\sub}&=\frac{M^{3/2} }{\pa{M^2-a^2}^{1/4}}\sqrt{\frac{\epsilon}{2}}\mathcal{A}_e,\\
    \mathcal{A}_e&=
    \begin{cases}
        e_i {\epsilon^{1-n_i}}, &0\leq n_i<1,e_i\neq0\\[1ex]
        \dfrac{aL_i}{M^2}+e_i, & n_i=1,e_i\neq0\\[2ex]
        \dfrac{aL_i}{M^2}, & (n_i>1,e_i\neq0) \text{ or }(e_i=0)
    \end{cases},
\end{align}
\end{subequations}
and
\begin{align}
    \label{eq:TSchwarz}
    T_{i\sub}&=\tilde{T}_{i\sub} + {ae_i}\pa{\dfrac{M^2ae_i}{4r_{\mathrm{H}}^2}-\dfrac{{L_i}}{2}}\delta_{n_i,0},\\
    \tilde{T}_{i\sub}&=Q_i+r_{\mathrm{H}}^2 \pa{1+\frac{L_i^2}{4M^2}}.
\end{align}
The functions $\tilde P_{i\sub}(L_i,e_i)$ and $T_{i\sub}(L_i,e_i)$ as written here apply for all spin. Finally, comparing $L_i(\mu_i,\ell_i)$~\eqref{eq:LScalingSchwarz} to $\mu_i(L_i,e_i)$~\eqref{eq:LScalingSubEK}, we find $e_i=2l_ir_\mathrm{H}/M$. By plugging $L_i(\mu_i,\ell_i)$ and $e_i=2\ell_ir_\mathrm{H}/M$ into $\tilde P_{i\sub}(L_i,e_i)$~\eqref{eq:PtildeSchwarz} and $T_{i\sub}(L_i,e_i)$~\eqref{eq:TSchwarz}, we recover $\tilde P_{i\sub}(\mu_i,\ell_i)$~\eqref{eq:PtildeSub} and $T_{i\sub}(\mu_i,\ell_i)$~\eqref{eq:TSub} for leading order in near-horizon scaling parameter $\epsilon$.

\section{High-energy collisions on the NHEK and near-NHEK backgrounds}
\label{app:NHEK}

This appendix reviews the near-horizon geometry of the extremal and near-extremal BHs---specifically the near-horizon extreme Kerr (NHEK) metric and near-NHEK metrics. We will then discuss collisions on these near-horizon backgrounds.

The near-horizon geometry of a (near-)extremal BH can be found by moving to another coordinate system ($\hat{x}^\nu=\cu{T,R,\theta,\Phi}$) defined through the coordinate transformation \cite{Bardeen1999,Hadar2014}
\begin{align}
    \label{eq:BardeenHorowitzCoords}
    t=\frac{2MT}{\kappa^p},\quad r=M\pa{1+R\kappa^p}, \quad\phi=\Phi+\frac{T}{\kappa^p},
\end{align}
where
\[\begin{alignedat}{2}
    \text{extremal BH:} &\quad a=M, &\quad & p=1, \\
    \text{near-extremal BH:} &\quad a=M\sqrt{1+\kappa^2}, &\quad & 0<p\leq1
\end{alignedat}\]
Applying this transformation to the Kerr metric \eqref{eq:Kerr},
and then taking a $\kappa\to 0$ limit, the Kerr metric becomes $d s^2=d {\hat{s}}^2+\O{\kappa^p}$
where the leading-order contribution is itself a solution to Einstein's equations. The limiting metric of the extremal BH is the NHEK metric. For the near-extremal BH the limiting metric is the NHEK metric when $0<p<1$. There is another limiting metric to the near-extremal BH when $p=1$, the near-NHEK metric.
These limiting metrics (NHEK and near-NHEK) take the form
\begin{align}
    \frac{d \hat s^2}{2M^2 \Gamma }=&-\pa{R^2-\sigma} \ed T^2+\frac{\ed R^2}{R^2-\sigma}+\ed\theta^2+\Lambda^2\pa{d\Phi+R\ed T}^2,
\end{align}
where $\Gamma=\pa{1+\cos^2\theta}/{2}$ and $\Lambda={2\sin\theta}/\pa{1+\cos^2\theta}$, and where 
\begin{align}
    \sigma=
    \begin{cases}
        0, & \text{NHEK}\\
        1, & \text{near-NHEK}
    \end{cases}.
\end{align}
The (near-)NHEK horizon is located at $R_\mathrm{H}=\sigma$. (Note: The near-NHEK metric is locally isomorphic to the NHEK metric. Despite this fact, each $p$ is physically meaningful since each (near-)NHEK describes distinct sets of radii. See Sec.~\ref{sec:NearHorizonCriticalLimit} for details.)

Timelike geodesic motion on the (near-)NHEK background can be parameterized by the particle mass $m$,
specific energy $\mathcal{U}$, 
specific azimuthal angular momentum $\mathcal{L}$, and specific Carter constant $\mathcal{Q}$:
\begin{subequations}
\begin{align}
    m&=\sqrt{-\hat{p}^\nu \hat{p}_\nu},\qquad
    \mathcal{U}=-\frac{\hat{p}_T}{m},\qquad
    \mathcal{L}=\frac{\hat{p}_\Phi}{m},\\
    \mathcal{Q}&=\frac{\hat{p}_\theta^2 +\pa{\Lambda^{-2}-1}\hat{p}_\Phi^2,}{m^2}+{2M^2\Gamma},\\
        &=-\frac{\hat{p}_r^2+\pa{\hat{p}_T-\hat{p}_\Phi R}^2}{m^2\pa{R^2-\sigma}}-\frac{{\hat{p}_\Phi}^2}{m^2}.
\end{align}
\label{eq:ConservedQuantitiesNHEK}
\end{subequations}
The momentum $\hat{p}=\hat{p}_\nu\ed\hat{x}^\mu$ is given by 
\begin{align}
    \label{eq:NHEKMomentum}
	\frac{\hat p}{m}=&-\mathcal{U}\ed T+\bar{s}_R\frac{\sqrt{{\mathbf{R}}(R)}}{R^2-\sigma}\ed R +\bar{s}_{\theta}\sqrt{\mathbf{\Theta}(\theta)}\ed\theta+\mathcal{L}\ed\Phi,
\end{align}
where the radial and angular potentials are
\begin{align}
    \label{eq:RadialPotentialNHEK}
    \mathbf{\Theta}&=\mathcal{Q}+\pa{1-\Lambda^{-2}}\mathcal{L}^2-{2M^2\Gamma},\\
    \label{eq:AngularPotentialNHEK}
    \mathbf{R}&=\mathcal{P}^2-\pa{R^2-\sigma}\mathcal{T},
\end{align}
with $\mathcal P=\mathcal{U}+\mathcal{L}R$ and $\mathcal{T}=\mathcal{Q}+\mathcal{L}^2$. Additionally,
\begin{subequations}
\label{eq:FourMomentum}
\begin{align}
    \frac{ \hat{p}^T}{m}&=\frac{\mathcal{P}}{2M^2\Gamma\pa{R^2-\sigma}},\\
    \frac{\hat{p}^R}{m}&=\bar{s}_r\frac{\sqrt{\bf{R}}}{2M^2\Gamma},\\
    \frac{\hat{p}^\theta}{m}&=\bar{s}_\theta\frac{\sqrt{\bf{\Theta}}}{2M^2\Gamma},\\
    \frac{\hat{p}^\Phi}{m}&=\frac{1}{2M^2\Gamma}\pa{\frac{\mathcal{L}}{\Lambda^2}-\frac{R\mathcal{P}}{R^2-\sigma}}.
\end{align}
\end{subequations}
The CM energy is then given by 
\begin{align}
    \frac{E_\cm^2}{m_1m_2}=&\frac{m_1^2+m_2^2}{m_1m_2}+\frac{{X-Y}}{M^2 \Gamma},\label{eq:EcmNHEKgeneral}\\
    X=&\frac{\mathcal{P}_1\mathcal{P}_2-\bar{s}_{R_1}\bar{s}_{R_2}\sqrt{\mathbf{R}_1}\sqrt{\mathbf{R}_2}}{R^2-\sigma},\label{eq:ECMchiNHEK}\\
    Y=&\frac{\mathcal{L}_1\mathcal{L}_2}{\Lambda^2}+ \bar{s}_{\theta_1}\bar{s}_{\theta_2} \sqrt{\mathbf\Theta_1}\sqrt{\mathbf\Theta_2}.
\end{align}

\subsection{near-horizon collisions}

We can examine the behavior of near-horizon collisions on the (near-)NHEK backgrounds. In the near-horizon limit of the CM energy \eqref{eq:EcmNHEK}, only the radial dependent term $X$ \eqref{eq:ECMchiNHEK} may be divergent. In the near-horizon limit $R\to \sigma$,
\begin{align}
    \label{eq:ECMchiNearHorizonNHEK}
    X=
    \begin{cases}
    \dfrac{{\mathcal{T}_2}\hat{\mathcal{U}}_1}{2\,\hat{\mathcal{U}}_2}+\dfrac{{\mathcal{T}_1}\hat{\mathcal{U}}_2}{2\,\hat{\mathcal{U}}_1}+\O{R-\sigma}, & s_R=1 \\[3ex]
    \dfrac{(1-s_R)\hat{\mathcal{U}}_1\hat{\mathcal{U}}_2}{\pa{1+\sigma}\pa{R-\sigma}^{2-\sigma}}+\O{\pa{R-\sigma}^{\sigma-1}}, & s_R\leq0
    \end{cases},
\end{align}
where $s_R\equiv\bar{s}_{R_1} \bar{s}_{R_2}$ and $\hat{\mathcal{U}}_i\equiv \mathcal{U}_i+\sigma \mathcal{L}_i$. Thus, criticality is defined by the specific energy
\begin{align}
    \mathcal{U}_*=-\sigma \mathcal{L}_i.
\end{align}
The behavior of the CM energy of collisions in the (near)-NHEK spacetimes mirrors that on the sub-extremal Kerr \eqref{eq:ECMchiNearHorizon} and extremal \eqref{eq:ECMchiNearHorizonEK} backgrounds: Types T and I collisions are manifestly divergent without tuning the conserved quantities of the particles, while Type V collisions are not. 
(Interestingly, for Types T and I collisions, $E_\cm^2 \sim \O{{R-R_\mathrm{H}}}$ on the sub-extremal and near-NHEK backgrounds and $E_\cm^2 \sim \O{\pa{R-R_\mathrm{H}}^{-2}}$ as in the extremal and near-extremal Kerr backgrounds.)

As in the other background examined, the near-horizon expression for $X$ \eqref{eq:ECMchiNearHorizonNHEK} is strictly applicable when the particles approach criticality slower than the collision radius approaches the horizon since it is calculated at fixed non-critical values of the specific energy. 
Rewriting $X$ \eqref{eq:ECMchiNHEK} as
\begin{align}
    \label{eq:X}
    X&=\tilde {\mathcal{P}}_1\tilde {\mathcal{P}}_2-s_{r}\sqrt{\tilde {\mathcal{P}}_1^2-{\mathcal{T}}_1}\sqrt{\tilde {\mathcal{P}}_2^2-{\mathcal{T}}_2},
\end{align}
where $\tilde{\mathcal{P}}={\mathcal{P}}/\sqrt{R^2-\sigma}$,
the (possible) divergence of the CM energy can be diagnosed by examining $\tilde {\mathcal{P}}_i$ for particles which approach criticality in the near-horizon limit. When $X$ is divergent, the CM energy is approximately
\begin{align}
    \label{eq:EcmNHEK}
    \frac{{E}_{\cm}^2}{m_1m_2}\approx\frac{2X}{M^2 \pa{1+\cos^2\theta}}.
\end{align}

To resolve the CM energy \eqref{eq:EcmNHEK} to leading order, we must resolve $R-\sigma$ and $\mathcal{U}-\sigma\mathcal{L}$ to leading order. Thus, we adopt the ansatz,
\begin{subequations}
\label{eq:ScalingLimitNHEKAnsatz}
\begin{alignat}{2}
    R &= \sigma + \epsilon,  &\quad & \epsilon\ll1, \\
    \mathcal{U}_i &= -\sigma \mathcal{L}_i + M e_i \epsilon^{n_i}, &\quad & 0\leq n_i.
    \label{eq:EScalingNHEK}
\end{alignat}
\end{subequations}
Eq.~\eqref{eq:EScalingNHEK} allows us to describe three types of particles: generic particles ($e_i\neq0$ and $n_i=0$), near-critical particles ($e_i\neq 0$ and $n_i>0$), and precisely critical particles ($e_i=0$). Using this ansatz, we can define the near-horizon metric, near-critical, near-horizon multi-scaling limit of a quantity $f$ as
\begin{align}
    \label{eq:ScalingLimitNHEK}
    f(R,\mathcal{U}_1,\mathcal{U}_2)\Big\vert_{\text{Eq.\eqref{eq:ScalingLimitNHEKAnsatz}}} \overset{\epsilon \to 0}{\sim} \ f_{\nhm}(\epsilon).
\end{align}

To leading order in the multi-scaling limit,
\begin{subequations}
\begin{align}
    \tilde{\mathcal{P}}_{i\nhm}=&M\pa{\frac{\epsilon}{2}}^{\sigma/2}\mathcal{A}_e,\\
    \mathcal{A}_e=&
    \begin{cases}
        \dfrac{e_i}{\epsilon^{1-n_i}},& n_i<1,e_i\neq0\\[2ex]
        \dfrac{\mathcal{L}_i}{M}+e_i, & n_i=1,e_i\neq0\\[2ex]
        \dfrac{\mathcal{L}_i}{M}, & (n_i>1,e_i\neq0) \text{ or }(e_i=0)
    \end{cases}.
\end{align}  
\end{subequations}
$\mathcal{T}_i$, independent of $\mathcal{U}_i$ and $R$, remains unchanged in the multi-scaling limit. Particle motion is only allowed when $\mathcal{R}_i\propto\tilde {\mathcal{P}}_i^2-\mathcal{T}_i\geq0$. $\mathcal{T}_i$ is finite and non-vanishing, so we must additionally restrict $n_i\leq1/2$ when $p=1$ to ensure that $\tilde{\mathcal{P}}_{i\nhm}$ is also non-vanishing

Divergent CM energy requires that at least one particle satisfies 
$$\text{divergence condition:}\ e_i\neq0,\ n_i<1.$$ 
For singly sourced divergent collisions of all Types, wherein particle 1 satisfies the divergence condition,
\begin{align}
    \label{eq:SinglySourcedNHEK}
    X_{\nhm}&=\tilde{\mathcal{P}}_{1\nhm}\tilde {\mathcal{P}}_{2\nhm}-s_{r}\sqrt{\tilde {\mathcal{P}}_{1\nhm}^2}\sqrt{\tilde {\mathcal{P}}_{2\nhm}^2-{\mathcal{T}}_{2\nhm}} \propto\frac{1}{\epsilon^{1-\sigma/2-n_1}}.
\end{align}
For doubly sourced divergence of Type V collisions, where $n_1\leq n_2$,
\begin{align}
    \label{eq:DoublySourcedTypeIandTNHEK}
    X_{\nhm}=(1-s_{r})\tilde{\mathcal{P}}_{1\nhm}\tilde {\mathcal{P}}_{2\nhm}\propto\frac{1}{\epsilon^{2-\sigma-n_2-n_1}}.
\end{align}
For doubly sourced divergent collisions of Type V, wherein $n_1<n_2$,
\begin{align}
    \label{eq:DoublySourcedTypeVNHEK}
    X_{\nhm}=\frac{{\mathcal{T}}_{2\nhm}{\tilde{\mathcal{P}}}_{1\nhm} }{2{\tilde{\mathcal{P}}}_{2\nhm}}\propto\frac{1}{\epsilon^{n_2-n_1}}.
\end{align}
The doubly divergent collision formulae Eqs.~\eqref{eq:DoublySourcedTypeIandTNHEK}-\eqref{eq:DoublySourcedTypeVNHEK} match the cases Eq.~\eqref{eq:ECMchiNearHorizonNHEK} when substituting $\mathcal{U}_i$ of the form \eqref{eq:EScalingNHEK}. The singly divergent collision formula Eq.~\eqref{eq:SinglySourcedNHEK} cannot be deduced from Eq.~\eqref{eq:ECMchiNearHorizonNHEK}. 

Among the equatorially bound orbits ($\theta=\pi/2,\ \hat{p}_\theta=0$,\ $\mathcal{Q}=M^2-3 \mathcal{L}^2/4$), (near-)NHEK metrics admit prograde circular orbits.
\begin{align}
    \mathcal{U}_\circ=-\sigma{\mathcal{L}_\circ},\quad \mathcal{L}_\circ= \dfrac{2MR}{\sqrt{3R^2-4 \sigma}},\quad R>\dfrac{2}{\sqrt{3}}\sigma, 
\end{align}
which are precisely critical particles.
On the NHEK background, the circular orbiters go down to the horizon, and can participate in a near-horizon Type T collisions. In contrast, on the near-NHEK background, circular orbits do not go to the horizon. But, as precisely critical particles, circular orbit plungers can participate in singly divergent Type V collision.

The near-horizon Type V collision on the NHEK background was first discussed by Galajinsky in Ref.~\cite{Galajinsky2013}. Therein it was noted that the divergent CM energy, near-horizon Type V collision on the extreme Kerr background found by BSW does not map onto a near-horizon Type V collision on the NHEK background. This discrepancy arises because the BSW analysis entails calculating the CM energy on the extreme Kerr background then taking the near-horizon limit, whereas the Galajinsky's analysis involves taking the near-horizon limit first to arrive at the NHEK background and then calculating the CM energy within that geometry. 
The order of limits is crucial: working only in terms of regular quantities in the (near-)NHEK geometry can restrict the physics under consideration, as regular quantities in the asymptotically flat region of the full BH geometry may not correspond to regular quantities in the near-horizon geometry.
As such, if our (near-)NHEK background is part of the throat of a (near-)extremal BH, high-energy collisions do not require the collision radius to be near the (near-)NHEK horizon.

\subsection{collisions with (near-)extremal Kerr particles}

If the near-horizon geometry (NHEK or near-NHEK) is embedded in the throat region of an asymptotically flat Kerr BH, we can use information inherited from the Kerr geometry in the near-horizon multi-scaling limits \eqref{eq:ScalingLimit} and \eqref{eq:ScalingLimitEK} to inform us about the (near-)NHEK conserved quantities for the particles \cite{Gralla2016, Gates2018}. 

Applying the coordinate transformation \eqref{eq:BardeenHorowitzCoords} to the Kerr momentum, we can define the Kerr specific energy and angular momentum in this coordinate system as
\begin{align}
    \label{eq:EnergyNHEKfromKerr}
    \bar{\mathcal{U}}&=-\frac{p_T}{m}=\frac{2M\mu-L}{\kappa^{p}},\\
    \bar{\mathcal{L}}&=\frac{p_\Phi}{m}= L.
\end{align}
Note that $\bar{\mathcal{U}}$ indicates that particles scaling into the near-horizon geometry can acquire divergent (near-)NHEK specific energy.
Consequently, they can seed high-energy collisions at any radius in the (near-)NHEK band.
Using the angular component $p_\theta$ and the functional forms of the Kerr and (near-)NHEK angular potentials, $\Theta$~\eqref{eq:AngularPotential} and $\mathbf{\Theta}$~\eqref{eq:AngularPotentialNHEK}, respectively:
\begin{gather}
    \frac{p_\theta}{m}= \bar{s}_\theta \sqrt{\mathbf{\Theta}(\bar{\mathcal{Q}},\bar{\mathcal{L}})} =\bar{s}_\theta \sqrt{\Theta(Q,L,\mu)},\\
    \big\Downarrow \nonumber\\
    \bar{\mathcal{Q}}= {2M^2\Gamma}-\pa{1-\Lambda^{-2}}\bar{\mathcal{L}}^2+\Theta(Q,L,\mu).
\end{gather}
(Note that $\bar{\mathcal{Q}}$ is not a conserved quantity of the Kerr geometry.) 

To map collisions from near-extremal or extremal BH backgrounds to those on the near-horizon backgrounds, we need ensure that the radial transformation in \eqref{eq:BardeenHorowitzCoords} matches Eq.~\eqref{eq:RScaling} or Eq.~\eqref{eq:RScalingEK} which impose $p=p_c$; and, additionally imposes
$ \kappa^p={\epsilon}/{R}$ with $ 0<p<1$ for the extremal BH. Setting 
\begin{align}
    ({\mathcal{U}}_{i},{\mathcal{L}}_{i},{\mathcal{Q}}_{i})&=\pa{\bar{\mathcal{U}}_{i},\bar{\mathcal{L}}_{i},\bar{\mathcal{Q}}_{i}},
\end{align}
we can analyze 
$E_{\cm}$ \eqref{eq:EcmNHEKgeneral} under the multi-scaling limits for the near-extremal BH \eqref{eq:ScalingLimit} and extremal BH \eqref{eq:ScalingLimitEK}.

Let us focus only on the high-energy collisions \eqref{eq:EcmNHEK}. The CM energy is divergent when the radially dependent CM energy term $X$~\eqref{eq:X} is divergent. Evaluated in the near-extremal multi-scaling limit,
\begin{align}
    \tilde{\mathcal{P}}_{i\near} = \dfrac{\pa{\bar{\mathcal{L}}_{i} +\bar{\mathcal{U}}_{i} R}_{\near}}{\sqrt{R^2-\delta_{p_c,1}}}= \tilde{P}_{i\near},
\end{align}
where $\delta_{x,y}$ is the Kronecker delta function and where $\tilde{P}_{i\near}$ is given by Eq.~\eqref{eq:TildesPnear} which, recall, is divergent when $l_i\neq0,\ q_i<p_c$. 
Additionally,
\begin{align}
    \tilde{\mathcal{T}}_{i\near} &=\pa{\bar{\mathcal{Q}}+\bar{\mathcal{L}}^2}_{\near}=\tilde{T}_{i\near}+M^2l_i(l_i-2\mu_i)\sin^2\theta\ \delta_{q_i,0},
\end{align}
where $\tilde{T}_{i\near}$ is given by Eq.~\eqref{eq:TildeTnear} which is remains finite. Hence, 
\begin{align}
    X_{\near}=\chi_{\near}\text{ for }l_1\neq0,\ q_1<p_c,
\end{align}
where $\chi_{\near}$ is given by Eqs.~\eqref{eq:SinglySourced}-\eqref{eq:DoublySourcedTypeV}. 
Thus, we recover the high-energy CM energy behavior  calculated in Sec.~\ref{sec:HighEnergyCollisions}.

Similarly, when evaluated in the extremal multi-scaling limit,
\begin{align}
    \tilde{\mathcal{P}}_{i\ext} &= \dfrac{\pa{\bar{\mathcal{L}}_{i} +\bar{\mathcal{U}}_{i} R}_{\ext}}{R}= \tilde{P}_{i\ext},
\end{align}
and
\begin{align}
    \tilde{\mathcal{T}}_{i\ext}&=\pa{\bar{\mathcal{Q}}_{i}+\bar{\mathcal{L}}_{i}^2}_{\ext}=\tilde{T}_{i\ext}+M^2\ell_i(\ell_i-2\mu_i)\sin^2\theta\ \delta_{n_i,0},
\end{align}
where $\tilde{\mathcal{P}}_{i\ext}$ is given by \eqref{eq:Pext} which only diverges when $\ell_i\neq0,\ n_i<1$, and where $\tilde{\mathcal{T}}_{i\ext}$ is given by \eqref{eq:Text} which is remains finite. Thus,
\begin{align}
    X_{\ext}=\chi_{\ext}\text{ for } \ell_1\neq0,\ n_1<1,
\end{align}
where $\chi_{\ext}$ is given by Eqs.~\eqref{eq:SinglySourcedExt}-\eqref{eq:DoublySourcedTypeVExt}. Hence, we recover the high-energy CM energy behavior calculated in App.~\ref{app:EKCMEnergy}.

\section{High-energy collisions with precisely critical particles on near-extremal Kerr background}
\label{app:PreciselyCritical}

\begin{figure}
    \centering
    \includegraphics[width=.5\linewidth]{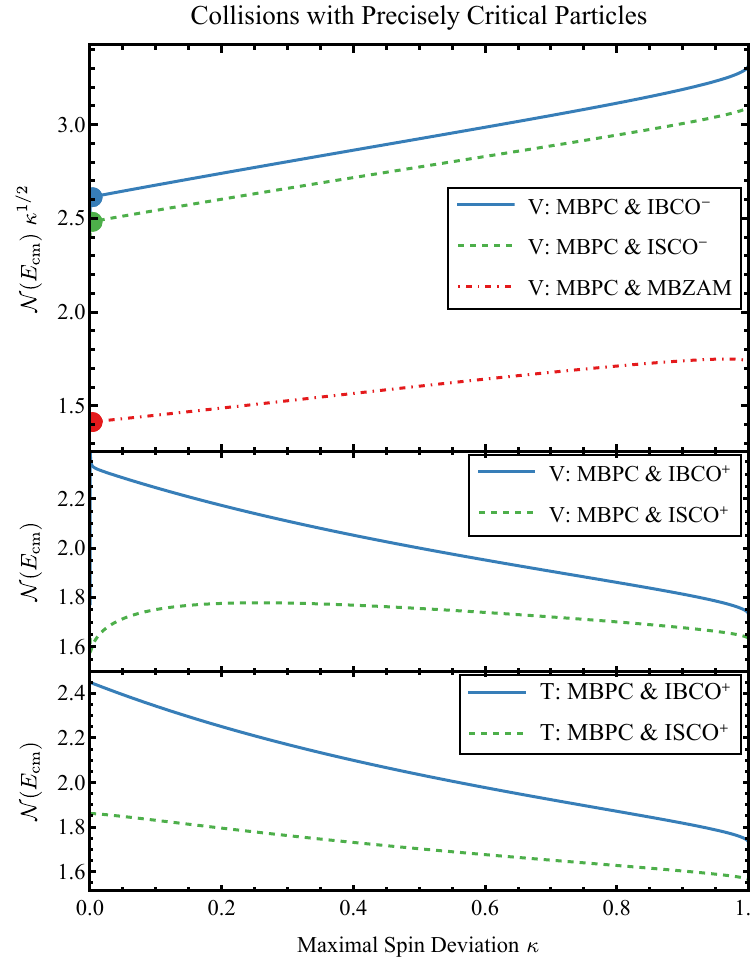}
    \caption{CM energies of collision including precisely critical particles. We plot the normalized CM energy $\mathcal{N}(E_{\cm})$ \eqref{eq:NCMEF} as a function of spin deviation parameter $\kappa=\sqrt{1-a^2/M^2}$.
    \\[1ex]
    {\bf Top panel:} Type V collisions at the horizon of equatorially bound, marginally bound precisely critical particles (labeled ``MBPC'') and generic particles: retrograde IBCO plunger (``IBCO$^-$''), ISCO plunger (``ISCO$^-$''), and marginally bound zero angular momentum particle (``MBZAM'').\\ 
    {\bf Middle panel:} Type V collisions at the horizon of equatorially bound, marginally bound precisely critical particles and prograde IBCO/ISCO plungers (``IBCO$^+$''/``ISCO$^+$'').\\
    {\bf Bottom panel:} Type T collision of equatorially bound, marginally bound precisely critical particles and prograde IBCO/ISCO orbiters (``IBCO$^+$''/``ISCO$^+$'').\\[1ex]
    In the near-horizon limit, the CM energy of the Type V collision of a precisely critical particle and a generic particle and the CM energy of the Type T collision of a circular orbiter and a generic particle  diverge ($E_{\cm}\sim\mathcal{O} (\kappa^{-1/2})$ for $\kappa\to0$). Thus, the normalized CM energy has been multiplied by $\kappa^{1/2}$ in the top panel to regularize the CM energy in the near-extremal limit $\kappa\to0$. 
    The analytic formula for the high-energy CM energy $E_{\cm \near}/\sqrt{m_1m_2}$ for the Type V collision of a precisely critical and generic particle is given by Eq.~\eqref{eq:TypeVPC}, and is illustrated as dots along the $\kappa=0$ axis in the top panel. The CM energy of a collision involving a near-horizon circular orbit particle and a precisely critical particle is finite in the near-horizon limit.
    }
    \label{fig:CollisionsPC}
\end{figure}

Here we consider collisions involving precisely critical particles around the near-extremal BH. A precisely critical particle does not meet the divergence condition, so the CM energy of any collision including a precisely critical particle is finite or singularly divergent.

In Fig.~\ref{fig:CollisionsPC}, we plot the CM energy of precisely critical particles with the particles described in Tab.~\ref{tab:NearHorizonNearCriticalParams}. Since near-horizon circular orbit particles ($l_1\neq0, q_2\geq p_c$) do not meet the divergence condition (see Sec.~\ref{sec:CircularOrbiterCollisions}), the near-horizon collision of precisely critical particle and near-horizon circular orbit particle has finite CM energy (see middle and bottom panels in Fig.~\ref{fig:CollisionsPC}). But, generic particles ($l_1\neq0, q_2=0$) do meet the divergence condition; and the CM energy of a precisely critical particle and a generic particle is divergent (see top panel in Fig.~\ref{fig:CollisionsPC}). 

The general formula for the CM energy of a singly sourced divergent Type V collision involving a precisely critical particle is 
\begin{subequations}
\label{eq:SinglyDivergentTypeVPlungePC}
\begin{align}
     \frac{E_\mathrm{cm\near}^2}{m_1m_2}=&\frac{4 l_1 \pa{2\mu_2-\sqrt{R^2\mathcal{J}+\mathcal{K}\delta_{p_c,1}}}}{\pa{1+\cos^2\theta}\pa{{R^2-\delta_{p_c,1}}}\kappa^{p_c-q_1}},\\
    \mathcal{J}&=3\mu_2^2-(1+\eta_2),\\
    \mathcal{K}&=1+\eta_2+\mu_2^2.
\end{align}
\end{subequations}

(We note that the $p_c<1$ case matches Eq.~\eqref{eq:SinglyDivergentTypeVPlunge2}, while the $p_c=1$ case matches Eq.~\eqref{eq:SinglyDivergentTypeVPlunge} with $q_2>1$. This agreement arises because $\tilde P_{i\near}$~\eqref{eq:TildesPnear} takes the same form for the precisely critical particle and near-critical particle with $q_i>p_c$.)

In particular, when a generic particle collides with a precisely critical particle that is equatorially bound ($\eta_2=0$) and marginally bound ($\mu_2=1$) at the horizon ($R=1, p_c=1$), the CM energy is
\begin{align}
    \label{eq:TypeVPC}
    \frac{E_{\cm\near}^2}{m_1m_2}=\frac{2 l_1}{\kappa}.
\end{align}
This contrasts with the CM energy of the analogous collision on the extremal BH background given by Eq.~\eqref{eq:TypeVPCExt}. Instead, near-horizon collisions on the extremal BH background map to collisions in the $0<p<1$ bands of the near-extremal BH background (see Sec.~\ref{app:EKCMEnergy}).

\bibliography{NHEKCollisions.bib}

\providecommand{\href}[2]{#2}\begingroup\raggedright\begin{thebibliography}{10}

\bibitem{Banados2009}
M.~{Ba{\~n}ados}, J.~{Silk}, and S.~M. {West}, ``{Kerr Black Holes as Particle
  Accelerators to Arbitrarily High Energy},''
  \href{http://dx.doi.org/10.1103/PhysRevLett.103.111102}{{\em \prl} {\bfseries
  103} no.~11, (Sept., 2009) 111102},
  \href{http://arxiv.org/abs/0909.0169}{{\ttfamily arXiv:0909.0169 [hep-ph]}}.

\bibitem{Jacobson2010}
T.~{Jacobson} and T.~P. {Sotiriou}, ``{Spinning Black Holes as Particle
  Accelerators},'' \href{http://dx.doi.org/10.1103/PhysRevLett.104.021101}{{\em
  \prl} {\bfseries 104} no.~2, (Jan., 2010) 021101},
  \href{http://arxiv.org/abs/0911.3363}{{\ttfamily arXiv:0911.3363 [gr-qc]}}.

\bibitem{Grib2010}
A.~A. {Grib} and Y.~V. {Pavlov}, ``{On particle collisions near Kerr's black
  holes},'' \href{http://dx.doi.org/10.48550/arXiv.1007.3222}{{\em arXiv
  e-prints} (July, 2010) arXiv:1007.3222},
  \href{http://arxiv.org/abs/1007.3222}{{\ttfamily arXiv:1007.3222 [gr-qc]}}.

\bibitem{Grib2011a}
A.~A. {Grib} and Y.~V. {Pavlov}, ``{On particle collisions in the gravitational
  field of the Kerr black hole},''
  \href{http://dx.doi.org/10.1016/j.astropartphys.2010.12.005}{{\em
  Astroparticle Physics} {\bfseries 34} no.~7, (Feb., 2011) 581--586},
  \href{http://arxiv.org/abs/1001.0756}{{\ttfamily arXiv:1001.0756 [gr-qc]}}.

\bibitem{Harada2011a}
T.~{Harada} and M.~{Kimura}, ``{Collision of an innermost stable circular orbit
  particle around a Kerr black hole},''
  \href{http://dx.doi.org/10.1103/PhysRevD.83.024002}{{\em \prd} {\bfseries 83}
  no.~2, (Jan., 2011) 024002}, \href{http://arxiv.org/abs/1010.0962}{{\ttfamily
  arXiv:1010.0962 [gr-qc]}}.

\bibitem{Harada2011b}
T.~{Harada} and M.~{Kimura}, ``{Collision of two general geodesic particles
  around a Kerr black hole},''
  \href{http://dx.doi.org/10.1103/PhysRevD.83.084041}{{\em \prd} {\bfseries 83}
  no.~8, (Apr., 2011) 084041}, \href{http://arxiv.org/abs/1102.3316}{{\ttfamily
  arXiv:1102.3316 [gr-qc]}}.

\bibitem{Bardeen1972}
J.~M. {Bardeen}, W.~H. {Press}, and S.~A. {Teukolsky}, ``{Rotating Black Holes:
  Locally Nonrotating Frames, Energy Extraction, and Scalar Synchrotron
  Radiation},'' \href{http://dx.doi.org/10.1086/151796}{{\em \apj} {\bfseries
  178} (Dec., 1972) 347--370}.

\bibitem{Bardeen1999}
J.~{Bardeen} and G.~T. {Horowitz}, ``{Extreme Kerr throat geometry: A vacuum
  analog of AdS$_{2}\times$S$^{2}$},''
  \href{http://dx.doi.org/10.1103/PhysRevD.60.104030}{{\em \prd} {\bfseries 60}
  no.~10, (Nov., 1999) 104030},
  \href{http://arxiv.org/abs/hep-th/9905099}{{\ttfamily arXiv:hep-th/9905099
  [hep-th]}}.

\bibitem{Kerr1963}
R.~P. Kerr, ``Gravitational field of a spinning mass as an example of
  algebraically special metrics,''
  \href{http://dx.doi.org/10.1103/PhysRevLett.11.237}{{\em Phys. Rev. Lett.}
  {\bfseries 11} (Sep, 1963) 237--238}.
  \url{https://link.aps.org/doi/10.1103/PhysRevLett.11.237}.

\bibitem{Boyer1967}
R.~H. {Boyer} and R.~W. {Lindquist}, ``{Maximal Analytic Extension of the Kerr
  Metric},'' \href{http://dx.doi.org/10.1063/1.1705193}{{\em Journal of
  Mathematical Physics} {\bfseries 8} no.~2, (Feb., 1967) 265--281}.

\bibitem{Carter:1968rr}
B.~Carter, ``{Global structure of the Kerr family of gravitational fields},''
  \href{http://dx.doi.org/10.1103/PhysRev.174.1559}{{\em Phys. Rev.} {\bfseries
  174} (1968) 1559--1571}.

\bibitem{Harada2014}
T.~{Harada} and M.~{Kimura}, ``{Black holes as particle accelerators: a brief
  review},'' \href{http://dx.doi.org/10.1088/0264-9381/31/24/243001}{{\em
  Classical and Quantum Gravity} {\bfseries 31} no.~24, (Dec., 2014) 243001},
  \href{http://arxiv.org/abs/1409.7502}{{\ttfamily arXiv:1409.7502 [gr-qc]}}.

\bibitem{Zaslavskii2012b}
O.~B. {Zaslavskii}, ``{Circular orbits and acceleration of particles by
  near-extremal dirty rotating black holes: general approach},''
  \href{http://dx.doi.org/10.1088/0264-9381/29/20/205004}{{\em Classical and
  Quantum Gravity} {\bfseries 29} no.~20, (Oct., 2012) 205004},
  \href{http://arxiv.org/abs/1201.5351}{{\ttfamily arXiv:1201.5351 [gr-qc]}}.

\bibitem{Grib2017}
A.~A. {Grib} and Y.~V. {Pavlov}, ``{Black holes and particles with zero or
  negative energy},'' \href{http://dx.doi.org/10.1134/S0040577917020088}{{\em
  Theoretical and Mathematical Physics} {\bfseries 190} no.~2, (Feb., 2017)
  268--278}, \href{http://arxiv.org/abs/1601.02592}{{\ttfamily arXiv:1601.02592
  [gr-qc]}}.

\bibitem{Wei2010}
S.-W. {Wei}, Y.-X. {Liu}, H.~{Guo}, and C.-E. {Fu}, ``{Charged spinning black
  holes as particle accelerators},''
  \href{http://dx.doi.org/10.1103/PhysRevD.82.103005}{{\em \prd} {\bfseries 82}
  no.~10, (Nov., 2010) 103005},
  \href{http://arxiv.org/abs/1006.1056}{{\ttfamily arXiv:1006.1056 [hep-th]}}.

\bibitem{Zaslavskii2010}
O.~B. {Zaslavskii}, ``{Acceleration of particles as a universal property of
  rotating black holes},''
  \href{http://dx.doi.org/10.1103/PhysRevD.82.083004}{{\em \prd} {\bfseries 82}
  no.~8, (Oct., 2010) 083004}, \href{http://arxiv.org/abs/1007.3678}{{\ttfamily
  arXiv:1007.3678 [gr-qc]}}.

\bibitem{Zaslavskii2012a}
O.~B. {Zaslavskii}, ``{Ultra-high energy collisions of nonequatorial geodesic
  particles near dirty black holes},''
  \href{http://dx.doi.org/10.1007/JHEP12(2012)032}{{\em Journal of High Energy
  Physics} {\bfseries 2012} (Dec., 2012) 32},
  \href{http://arxiv.org/abs/1209.4987}{{\ttfamily arXiv:1209.4987 [gr-qc]}}.

\bibitem{Bredberg2010}
I.~{Bredberg}, T.~{Hartman}, W.~{Song}, and A.~{Strominger}, ``{Black hole
  superradiance from Kerr/CFT},''
  \href{http://dx.doi.org/10.1007/JHEP04(2010)019}{{\em Journal of High Energy
  Physics} {\bfseries 2010} (Apr., 2010) 19},
  \href{http://arxiv.org/abs/0907.3477}{{\ttfamily arXiv:0907.3477 [hep-th]}}.

\bibitem{Gates2020}
D.~E.~A. {Gates}, S.~{Hadar}, and A.~{Lupsasca}, ``{Maximum observable
  blueshift from circular equatorial Kerr orbiters},''
  \href{http://dx.doi.org/10.1103/PhysRevD.102.104041}{{\em \prd} {\bfseries
  102} no.~10, (Nov., 2020) 104041},
  \href{http://arxiv.org/abs/2009.03310}{{\ttfamily arXiv:2009.03310 [gr-qc]}}.

\bibitem{Hod2017}
S.~{Hod}, ``{Marginally bound (critical) geodesics of rapidly rotating black
  holes},'' \href{http://dx.doi.org/10.48550/arXiv.1707.05680}{{\em arXiv
  e-prints} (July, 2017) arXiv:1707.05680},
  \href{http://arxiv.org/abs/1707.05680}{{\ttfamily arXiv:1707.05680 [gr-qc]}}.

\bibitem{Stein2020}
L.~C. {Stein} and N.~{Warburton}, ``{Location of the last stable orbit in Kerr
  spacetime},'' \href{http://dx.doi.org/10.1103/PhysRevD.101.064007}{{\em \prd}
  {\bfseries 101} no.~6, (Mar., 2020) 064007},
  \href{http://arxiv.org/abs/1912.07609}{{\ttfamily arXiv:1912.07609 [gr-qc]}}.

\bibitem{Teo2021}
E.~{Teo}, ``{Spherical orbits around a Kerr black hole},''
  \href{http://dx.doi.org/10.1007/s10714-020-02782-z}{{\em General Relativity
  and Gravitation} {\bfseries 53} no.~1, (Jan., 2021) 10},
  \href{http://arxiv.org/abs/2007.04022}{{\ttfamily arXiv:2007.04022 [gr-qc]}}.

\bibitem{Compere2021}
G.~Comp\`ere, Y.~Liu, and J.~Long, ``Classification of radial kerr geodesic
  motion,'' \href{http://dx.doi.org/10.1103/PhysRevD.105.024075}{{\em Phys.
  Rev. D} {\bfseries 105} (Jan, 2022) 024075}.
  \url{https://link.aps.org/doi/10.1103/PhysRevD.105.024075}.

\bibitem{Levin2009}
J.~{Levin} and G.~{Perez-Giz}, ``{Homoclinic orbits around spinning black
  holes. I. Exact solution for the Kerr separatrix},''
  \href{http://dx.doi.org/10.1103/PhysRevD.79.124013}{{\em \prd} {\bfseries 79}
  no.~12, (June, 2009) 124013},
  \href{http://arxiv.org/abs/0811.3814}{{\ttfamily arXiv:0811.3814 [gr-qc]}}.

\bibitem{Reynolds2019}
C.~S. {Reynolds}, ``{Observing black holes spin},''
  \href{http://dx.doi.org/10.1038/s41550-018-0665-z}{{\em Nature Astronomy}
  {\bfseries 3} (Jan., 2019) 41--47},
  \href{http://arxiv.org/abs/1903.11704}{{\ttfamily arXiv:1903.11704
  [astro-ph.HE]}}.

\bibitem{Draghis2023}
P.~A. Draghis, J.~M. Miller, M.~C. Brumback, A.~C. Fabian, J.~A. Tomsick, and
  A.~Zoghbi, ``{An Extreme Black Hole in the Recurrent X-ray Transient XTE
  J2012+381},'' \href{http://arxiv.org/abs/2307.06988}{{\ttfamily
  arXiv:2307.06988 [astro-ph.HE]}}.

\bibitem{Thorne1974}
K.~S. {Thorne}, ``{Disk-Accretion onto a Black Hole. II. Evolution of the
  Hole},'' \href{http://dx.doi.org/10.1086/152991}{{\em \apj} {\bfseries 191}
  (July, 1974) 507--520}.

\bibitem{Berti2009}
E.~{Berti}, V.~{Cardoso}, L.~{Gualtieri}, F.~{Pretorius}, and U.~{Sperhake},
  ``{Comment on ``Kerr Black Holes as Particle Accelerators to Arbitrarily High
  Energy''},'' \href{http://dx.doi.org/10.1103/PhysRevLett.103.239001}{{\em
  \prl} {\bfseries 103} no.~23, (Dec., 2009) 239001},
  \href{http://arxiv.org/abs/0911.2243}{{\ttfamily arXiv:0911.2243 [gr-qc]}}.

\bibitem{Novikov1973}
I.~D. {Novikov} and K.~S. {Thorne}, ``{Astrophysics of black holes.},'' in {\em
  Black Holes (Les Astres Occlus)}, pp.~343--450.
\newblock Jan., 1973.

\bibitem{Cunningham1975}
C.~T. {Cunningham}, ``{The effects of redshifts and focusing on the spectrum of
  an accretion disk around a Kerr black hole.},''
  \href{http://dx.doi.org/10.1086/154033}{{\em \apj} {\bfseries 202} (Dec.,
  1975) 788--802}.

\bibitem{Zhang2024}
W.~Zhang, M.~Dov\v{c}iak, M.~Bursa, J.~Svoboda, and V.~Karas, ``{Inferring the
  iron K emissivity profiles of accretion discs irradiated by extended
  coronae},'' \href{http://dx.doi.org/10.1093/mnras/stae1714}{{\em Mon. Not.
  Roy. Astron. Soc.} {\bfseries 532} no.~4, (2024) 3786--3796},
  \href{http://arxiv.org/abs/2407.08336}{{\ttfamily arXiv:2407.08336
  [astro-ph.HE]}}.

\bibitem{Brenneman2006}
L.~W. {Brenneman} and C.~S. {Reynolds}, ``{Constraining Black Hole Spin via
  X-Ray Spectroscopy},'' \href{http://dx.doi.org/10.1086/508146}{{\em \apj}
  {\bfseries 652} no.~2, (Dec., 2006) 1028--1043},
  \href{http://arxiv.org/abs/astro-ph/0608502}{{\ttfamily
  arXiv:astro-ph/0608502 [astro-ph]}}.

\bibitem{Fabian2000}
A.~C. Fabian, K.~Iwasawa, C.~S. Reynolds, and A.~J. Young, ``{Broad iron lines
  in active galactic nuclei},'' \href{http://dx.doi.org/10.1086/316610}{{\em
  Publ. Astron. Soc. Pac.} {\bfseries 112} (2000) 1145},
  \href{http://arxiv.org/abs/astro-ph/0004366}{{\ttfamily
  arXiv:astro-ph/0004366}}.

\bibitem{Reynolds2020}
C.~S. Reynolds, ``{Observational Constraints on Black Hole Spin},''
  \href{http://dx.doi.org/10.1146/annurev-astro-112420-035022}{{\em Ann. Rev.
  Astron. Astrophys.} {\bfseries 59} (2021) 117--154},
  \href{http://arxiv.org/abs/2011.08948}{{\ttfamily arXiv:2011.08948
  [astro-ph.HE]}}.

\bibitem{Hadar2014}
S.~{Hadar}, A.~P. {Porfyriadis}, and A.~{Strominger}, ``{Gravity waves from
  extreme-mass-ratio plunges into Kerr black holes},''
  \href{http://dx.doi.org/10.1103/PhysRevD.90.064045}{{\em \prd} {\bfseries 90}
  no.~6, (Sept., 2014) 064045},
  \href{http://arxiv.org/abs/1403.2797}{{\ttfamily arXiv:1403.2797 [hep-th]}}.

\bibitem{Hadar2017}
S.~{Hadar} and A.~P. {Porfyriadis}, ``{Whirling orbits around twirling black
  holes from conformal symmetry},''
  \href{http://dx.doi.org/10.1007/JHEP03(2017)014}{{\em Journal of High Energy
  Physics} {\bfseries 2017} no.~3, (Mar., 2017) 14},
  \href{http://arxiv.org/abs/1611.09834}{{\ttfamily arXiv:1611.09834
  [hep-th]}}.

\bibitem{Kapec2020}
D.~{Kapec} and A.~{Lupsasca}, ``{Particle motion near high-spin black holes},''
  \href{http://dx.doi.org/10.1088/1361-6382/ab519e}{{\em Classical and Quantum
  Gravity} {\bfseries 37} no.~1, (Jan., 2020) 015006},
  \href{http://arxiv.org/abs/1905.11406}{{\ttfamily arXiv:1905.11406
  [hep-th]}}.

\bibitem{Compere2020}
G.~{Comp{\`e}re} and A.~{Druart}, ``{Near-horizon geodesics of high spin black
  holes},'' \href{http://dx.doi.org/10.1103/PhysRevD.101.084042}{{\em \prd}
  {\bfseries 101} no.~8, (Apr., 2020) 084042},
  \href{http://arxiv.org/abs/2001.03478}{{\ttfamily arXiv:2001.03478 [gr-qc]}}.

\bibitem{Gates2023}
D.~E.~A. {Gates} and S.~{Hadar}, ``{Signatures of particle collisions near
  extreme black holes},''
  \href{http://dx.doi.org/10.48550/arXiv.2309.04572}{{\em arXiv e-prints}
  (Sept., 2023) arXiv:2309.04572},
  \href{http://arxiv.org/abs/2309.04572}{{\ttfamily arXiv:2309.04572 [gr-qc]}}.

\bibitem{Piran1975}
T.~{Piran}, J.~{Shaham}, and J.~{Katz}, ``{High Efficiency of the Penrose
  Mechanism for Particle Collisions},''
  \href{http://dx.doi.org/10.1086/181755}{{\em \apjl} {\bfseries 196} (Mar.,
  1975) L107}.

\bibitem{Piran1977}
T.~{Piran} and J.~{Shaham}, ``{Production of gamma-ray bursts near rapid
  rotating accreting black holes.},''
  \href{http://dx.doi.org/10.1086/155251}{{\em \apj} {\bfseries 214} (May,
  1977) 268--299}.

\bibitem{Piran1977a}
T.~{Piran} and J.~{Shaham}, ``{Upper bounds on collisional Penrose processes
  near rotating black-hole horizons},''
  \href{http://dx.doi.org/10.1103/PhysRevD.16.1615}{{\em \prd} {\bfseries 16}
  no.~6, (Sept., 1977) 1615--1635}.

\bibitem{Bejger2012}
M.~{Bejger}, T.~{Piran}, M.~{Abramowicz}, and F.~{H{\r{a}}kanson},
  ``{Collisional Penrose Process near the Horizon of Extreme Kerr Black
  Holes},'' \href{http://dx.doi.org/10.1103/PhysRevLett.109.121101}{{\em \prl}
  {\bfseries 109} no.~12, (Sept., 2012) 121101},
  \href{http://arxiv.org/abs/1205.4350}{{\ttfamily arXiv:1205.4350
  [astro-ph.HE]}}.

\bibitem{Leiderschneider2016}
E.~{Leiderschneider} and T.~{Piran}, ``{Maximal efficiency of the collisional
  Penrose process},'' \href{http://dx.doi.org/10.1103/PhysRevD.93.043015}{{\em
  \prd} {\bfseries 93} no.~4, (Feb., 2016) 043015},
  \href{http://arxiv.org/abs/1510.06764}{{\ttfamily arXiv:1510.06764 [gr-qc]}}.

\bibitem{Liberati2022}
S.~{Liberati}, C.~{Pfeifer}, and J.~{Relancio}, ``{Exploring black holes as
  particle accelerators: hoop-radius, target particles and escaping
  conditions},'' \href{http://dx.doi.org/10.1088/1475-7516/2022/05/023}{{\em
  \jcap} {\bfseries 2022} no.~5, (May, 2022) 023},
  \href{http://arxiv.org/abs/2106.01385}{{\ttfamily arXiv:2106.01385 [gr-qc]}}.

\bibitem{Galajinsky2013}
A.~{Galajinsky}, ``{Near horizon geometry of extremal black holes and
  Banados-Silk-West effect},''
  \href{http://dx.doi.org/10.48550/arXiv.1301.1159}{{\em arXiv e-prints} (Jan.,
  2013) arXiv:1301.1159}, \href{http://arxiv.org/abs/1301.1159}{{\ttfamily
  arXiv:1301.1159 [gr-qc]}}.

\bibitem{Gralla2016}
S.~E. {Gralla}, A.~{Lupsasca}, and A.~{Strominger}, ``{Near-horizon Kerr
  magnetosphere},'' \href{http://dx.doi.org/10.1103/PhysRevD.93.104041}{{\em
  \prd} {\bfseries 93} no.~10, (May, 2016) 104041},
  \href{http://arxiv.org/abs/1602.01833}{{\ttfamily arXiv:1602.01833
  [hep-th]}}.

\bibitem{Gates2018}
D.~{Gates}, D.~{Kapec}, A.~{Lupsasca}, Y.~{Shi}, and A.~{Strominger},
  ``{Polarization Whorls from M87 at the Event Horizon Telescope},''
  \href{http://dx.doi.org/10.1098/rspa.2019.0618}{{\em Proceedings of the Royal
  Society A} {\bfseries 476} no.~2237, (Sep, 2020) 20190618},
  \href{http://arxiv.org/abs/1809.09092}{{\ttfamily arXiv:1809.09092
  [hep-th]}}.

\end{thebibliography}\endgroup
\bibliographystyle{utphys2}

\end{document}